\documentclass[lettersize,journal]{IEEEtran}
\IEEEoverridecommandlockouts
\usepackage{stfloats}
\usepackage[english]{babel}
\usepackage{cite}
\usepackage{amsmath,amssymb,amsfonts}
\usepackage{algorithmic}
\usepackage{amsfonts}
\usepackage{graphicx}
\usepackage{epstopdf}
\usepackage{subfigure}
\usepackage[ruled,linesnumbered]{algorithm2e}
\usepackage{amsmath}
\usepackage{textcomp}
\usepackage{xcolor}
\usepackage{amssymb,amsmath}
\usepackage{enumitem}
\usepackage{pifont}
\usepackage{bm}
\usepackage{amsmath,lipsum}
\usepackage{cuted}
\usepackage{booktabs}
\usepackage[justification=centering]{caption}
\usepackage{mathtools}
\usepackage{tablefootnote}
\usepackage{amsthm}
\newlist{dblparenenum}{enumerate}{1}
\setlist[dblparenenum,1]{label={{(}\arabic*{{)}}}}
\newtheoremstyle{mystyle}               
  {0pt}                                 
  {0pt}                                 
  {\normalfont}                         
  {\parindent}                          
  {\itshape}                            
  {:}                                   
  { }                                   
  {\thmname{#1}\thmnumber{ #2}\thmnote{ (#3)}}  
\theoremstyle{mystyle}
\newtheorem{remark}{Remark}

\newtheorem{proposition}{Proposition}
\newenvironment{Proof}{{\noindent \textit {Proof:}}}{\hfill $\square$ \par}

\def\BibTeX{{\rm B\kern-.05em{\sc i\kern-.025em b}\kern-.08em
    T\kern-.1667em\lower.7ex\hbox{E}\kern-.125emX}}

\begin{document}

\title{Confidence Based Asynchronous Integrated Communication and Localization Networks Using Pulsed UWB Signals}
\author{Fan Liu, {\it Graduate Student Member, IEEE}, Bofeng Zheng, Tingting Zhang, {\it Member, IEEE}, \\ and Qinyu Zhang, {\it Senior Member, IEEE}
\thanks{
Fan Liu, Bofeng Zheng, Tingting Zhang and Qinyu Zhang are with the School of Electronics and Information Engineering, Harbin Institute of Technology, Shenzhen 518055, China. Fan Liu and Tingting Zhang are also with the Guangdong Provincial Key Laboratory of Space-Aerial Networking and Intelligent Sensing, Shenzhen 518055, China. Tingting Zhang and Qinyu Zhang are also with the Pengcheng Laboratory (PCL), Shenzhen 518055, China. (liufan0613@stu.hit.edu.cn; zbf031018@163.com; zhangtt@hit.edu.cn; zqy@hit.edu.cn).
}
}

\maketitle

\begin{abstract}
In recent years, UWB has garnered widespread attention in academia and industry due to its low power consumption, wide bandwidth, and high time resolution characteristics. This paper introduces the design of an asynchronous IR-UWB integrated communication and localization (ICL) downlink network, which employs unified waveforms to enable simultaneous data transmission and localization.
A differential sequential detection strategy has been proposed for data demodulation. To address errors caused by symbol misalignment, a novel symbol confidence metric model is introduced to ensure reliable pulse detection and time-of-arrival (TOA) estimation. Additionally, an asynchronous start-of-frame delimiter (SFD) detection model has been constructed to guide parameter optimization for practical applications.
Furthermore, the clock drift estimation has been improved by leveraging the confidence metric within a modified weighted least squares (MWLS) framework. Simulation results demonstrate that the proposed system achieves reliable clock drift estimation, communication, and self-localization simultaneously. The operational range of the confidence metric required for these outcomes is also quantified, providing valuable insights for parameter design and system implementation. Finally, the agent localization accuracy can be achieved within 10 cm at over 90\% confidence, with commercial UWB devices according to practical measurements.
\end{abstract}


\section{Introduction}

\IEEEPARstart{I}{mpulse} radio ultra-wideband (IR-UWB) is a promising technology that employs nanosecond-level impulses to transmit information, offering exceptionally high time resolution, low transmission power, and broad bandwidth. These characteristics make IR-UWB highly advantageous in wireless communications, ranging, sensing, detection, and localization applications \cite{UltraNF,ranging2009,UWBsurvey,ir_how_it_works}. In recent years, there has been substantial industrial interest in UWB, with applications expanding to smart home systems, mobile payment and access control systems, and medical monitoring devices \cite{Review,Survey2024}.
A notable development is the introduction of the IEEE 802.15.4 UWB Next Generation standard, proposed by the IEEE 802.15 Wireless Sensor Network Group 4ab (WSN TG4ab) \cite{IEEE4ab}, indicating the broad application potential of UWB technology.

The integration of communication and sensing(ISAC) or integration of communication and localization(ICL) has gained significant attention to improve spectrum and hardware efficiency \cite{JCLmm,JointSurvey3,JRCTcom}. In modern low-power systems, IR-UWB stands out as a competitive solution for ISAC applications due to its capability to achieve reliable data demodulation and range estimation, both heavily rely on precise time of arrival (TOA) estimations of pulses \cite{TR,SurveyMagzine,ISACVitalSign}.
We have developed a theoretical framework in terms of equivalent Fisher information matrix (EFIM) to characterize the coupling relationship between communication and sensing parameters within an IR-UWB ISAC system, along with decoupling strategies using pilot signals or differential techniques \cite{Zenan,XunZe,JingWen,liu2024fundament}. The feasibility of integrating communication and sensing through UWB signals has been well demonstrated.

With the increasing demand for multi-anchor ad-hoc networks, multiple emitters frequently operate simultaneously in the same radio environment, introducing multiuser interference (MUI) that significantly degrades TOA or angel of arrival (AOA) estimation \cite{Selflocation,ISACDownlink,JPBA,MultiUser}. IR-UWB systems address this challenge by utilizing known time-hopping (TH) codes \cite{THwin,TH2} or direct sequence (DS) spreading patterns \cite{DSSS,DSSS2}. Despite these efforts, accurate TOA estimation in complex, noisy environments remains a challenging task. Traditional "threshold-based" strategies often rely on hard decisions, which are prone to errors in the presence of dense multipath propagation and noise \cite{MILCOM,HumanTrack,Shen}. To overcome these limitations, recent studies have refined ranging and localization methods leveraging {\it soft information}, which provides richer insights compared to conventional single-value estimates and demonstrates superior performance in harsh wireless environments \cite{SoftRange,SoftLocal,FanTWC}.

Devices operating asynchronously experience clock offsets and drifts, leading to misalignments and making precise demodulation and ranging difficult. Most existing UWB signal processing studies assume synchronization between transmitted and received signals \cite{UWBOOK,UWBRIS}. Alternatively, synchronization is achieved through methods such as time of flight (TOF) or two way ranging (TWR), enabling data demodulation to proceed under synchronized conditions \cite{UWBSynch,UWBSynch2,UWBDrift,TimingSynchron}. These approaches typically focus on compensating for timing errors through estimation and correction algorithms. However, research on demodulating signals in the presence of clock asynchrony remains relatively scarce.

Pulse position modulation (PPM) is widely used in UWB systems as it simplifies transceiver design by avoiding the need for pulse inversion \cite{PPMADC,PPMRF}.
A notable shared feature of TH and PPM is pulses in the time domain, making them well-suited for integration into TH-PPM \cite{THPPM,THPPMScholtz}. In this paper, we aim to develop an integrated communication and localization (ICL) system using TH-PPM schemes in a multi-anchor network.
A downlink TDOA strategy is preferred, which in theory can accommodate infinite users. However, due to the unideal clock and propagation environments, the agent simultaneously performs clock drift estimation, demodulation, and self-localization, all while accounting for clock offsets between the anchors and the agent.

The key contributions of this work are summarized as follows:
\begin{dblparenenum}

\item A low-complexity demodulation solution based on differential sequential estimation is proposed for downlink multi-access environments, enabling simultaneous data symbol demodulation under asynchronous conditions, clock drift estimation, and pseudo-TOA estimation with code division. Moreover, this method is adaptable to multi-agent scenarios.

\item We propose a symbol confidence metric model, designed to mitigate error propagation in sequential detection processes. By integrating the confidence metrics derived from the symbol interval and the symbol amplitude, we develop a unified symbol confidence and establish a confidence area that simultaneously achieves optimal performance for both communication and sensing.

\item The symbol confidence is utilized as a weighting factor in the clock drift, where a modified weighted least squares (MWLS) algorithm has been employed to enhance the accuracy of the estimation process.

\item A downlink  time difference of arrival (TDOA)-based localization system is implemented utilizing commercially available UWB anchors, enabling joint clock calibration and accurate dynamic localization of the agent.\footnote {A demo video of the prototype system can be found at the attached supplementary file, or on line at: https://youtu.be/9HyMu1AsC5k.}

\end{dblparenenum}

The rest of the paper is organized as follows. Section \ref{SecModel} presents the IR-UWB based ICL system model with clock offset and drift.
Section \ref{ICLProcess} gives the ICL implementation process based on differential symbol detection strategy. SI-based symbol confidence metric models are presented in Section \ref{sec_soft} to realize the clock drift estimation and symbol detection enhancement. Finally, simulation and experiment results are given in Section \ref{section Numerical},  and conclusions are drawn in the last section.

\section{System Model}    \label{SecModel}

\subsection{Network Settings}

Consider a location-aware network of $N_\text{a}$ agent whose position is to be estimated and $N_\text{b}$ asynchronous pre-deployed anchors with known positions. Different values of asynchronous clock offset and drift exist between agents and anchors as depicted in Sec.~\ref{ClockModel}. The sets of anchors and agents can be expressed as $\mathcal{N}_\text{b}$ and $\mathcal{N}_\text{a}$, respectively. At the same time, the anchor performs data transmission to the agent. All agents could receive the IR-UWB based ICL signals {\it broadcasted} from all anchors, then perform data demodulation and self-localization.

\begin{figure}[h]
\centerline{\includegraphics[width = 0.8\columnwidth]{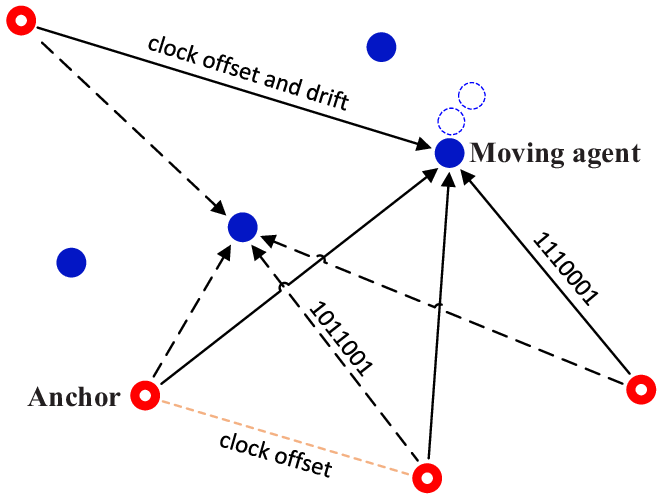}}
\caption{ICL network structure.}      \label{Network}
\end{figure}

The positions of the agents and anchors are expressed as $\mathbf{p}_n = {\big[\:{p_{\text x}^n},{p_{\text y}^n},{p_{\text z}^n} \: \big]^\text{T}}$ and
$\mathbf{r}_m = {\big[\: {x_m},{y_m},{z_m}\:\big]^\text{T}}$, respectively, where $n\in\mathcal{N}_\text{a}$, $m\in\mathcal{N}_\text{b}$.
The delay of signal propagation in the air from the $n$-agent to $m$-anchor is
\begin{equation}
\tau^{(mn)} = \frac{1}{c}{\left\| {{\mathbf{r}_m} - \mathbf{p}_n} \right\|_2},
\end{equation}
where $c = 3 \times 10^8 \:\: m/s$ is the speed of electromagnetic waves in the air, and ${\left\| \cdot \right\|_2}$ indicates the 2-norm.

\subsection{Clock Model}  \label{ClockModel}

Generally, there are two main issues, namely clock offset and drift, to be considered during clock synchronization
The clock model can be established to relate the local time of an agent to the reference anchor, which is described by the following formula:
\begin{equation}  \label{EqClock}
t_{\rm{c}}^{(n)} = \underbrace {\Delta {t^{(mn)}}}_{{\rm{offset}}} + \underbrace {{\varepsilon ^{(mn)}}t}_{{\rm{drift}}},
\end{equation}
where $t^{(n)}_{\text c}$ is the local time of agent $n$, $t$ is the reference time,
${\Delta t^{(mn)}}$ and $\varepsilon^{\text(mn)}t$ denote the initial clock offset and the clock drift between the agent $n$ and anchor $m$.

\begin{figure}  [h]
  \centering
  \includegraphics[width = 0.7\columnwidth]{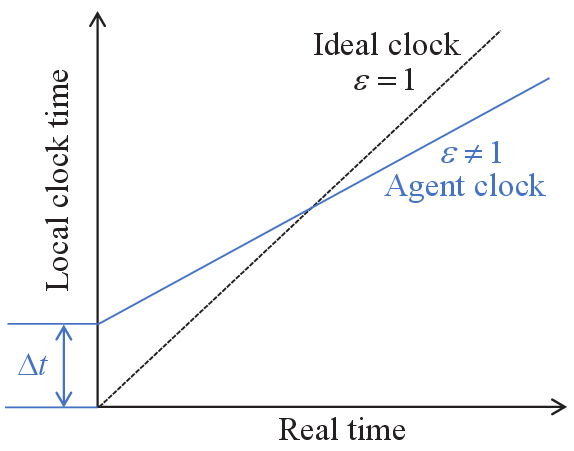}
  \caption{Clock model.}     \label{FigClock}
\end{figure}

The clock model is demonstrated as shown in Fig.~\ref{FigClock}. The slope of the ideal reference clock is 1. In a real clock system, the slope of the clock drift may deviate.

\subsection{Signal Model}

\begin{figure*} [b]
\hrulefill
\begin{equation}    \label{eqRXAgent}
{r^{(n)}}(t)  =  \sum\limits_{\kappa = 0}^{{N_\text f-1}} \sum\limits_{m = 1}^{{N_\text b}} {{\alpha^{(mn)}}{{s}^{(m)}} \Big(t - \Delta t^{(mn)} - \varepsilon^{(mn)} t \Big)}*h\Big( t-\tau^{(mn)}_{\kappa}  \Big) + z(t), {n \in\mathcal{N}_\text{a}},{m \in\mathcal{N}_\text{b}}
\end{equation}
\end{figure*}

\begin{figure*}[t]
\centerline{\includegraphics[width = 2.0\columnwidth]{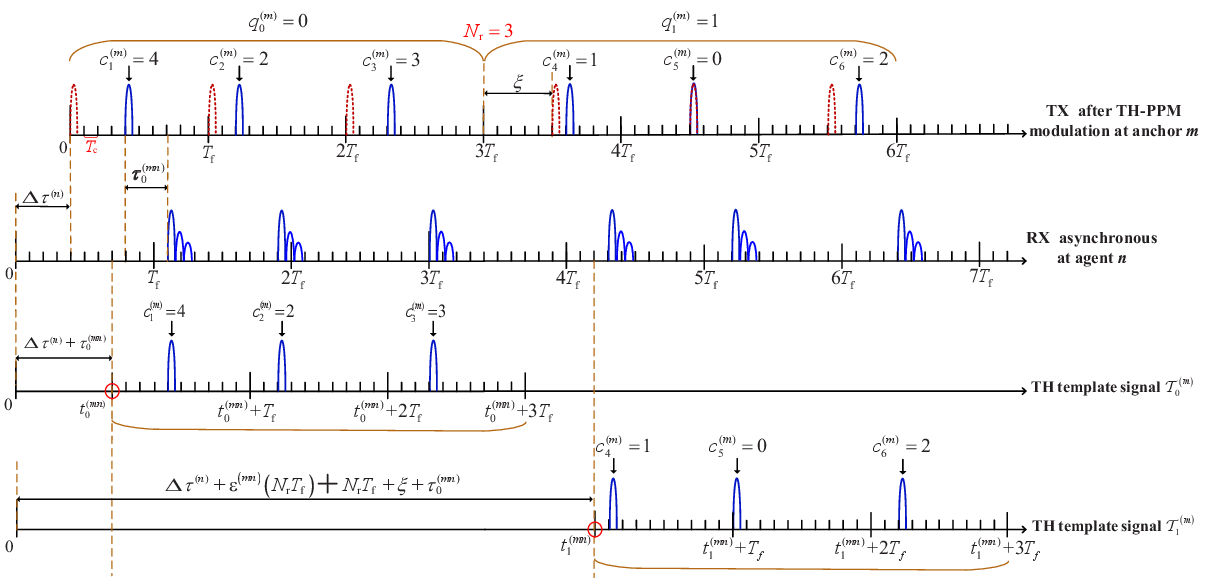}}
\caption{Schematic diagram of signal propagation delay of $(m,n)$-th link, $N_\text r=3$.} \label{Signal Propagation Delay}
\end{figure*}

Consider the proposed multiple access system with $N_\text{b}$ active UWB anchors, a typical TH format for the transmitted PPM signal of anchor $m$ is
\begin{equation}    \label{eq st}
\small
\begin{aligned}
s^{(m)}(t) = & \! \sqrt{{E_\text{tb}}}\sum\limits_{\kappa = 0}^{{N_{\text f}}-1}\sum\limits_{r = 0}^{{N_\text r}-1} \! {w\Big(t - {\kappa}{N_\text r}{T_f} - r{T_f}-{c^{(m)}_{\tilde \kappa}}{T_\text c}-{\xi} {q^{(m)}_\kappa} \Big)},   \\
& {m \in\mathcal{N}_\text{b}}
\end{aligned}
\end{equation}
where ${T_{\text f}}$ is bit repetition interval (PRI), ${E_\text{tb}}$ is the transmitted energy for one pulse, $N_{\text f}$ is the total number of data symbol to be transmitted, ${N_\text r}$ represents the number of repetition count of one data bit, $\tilde \kappa= {\kappa{N_\text r}+r}$.
$\xi$ is the time-shift of PPM and ${q^{(m)}_\kappa} \in \left\{ {0,1} \right\}$ represents the $\kappa$-th data symbol to be transmitted. Each agent is assigned a distinctive time-shift pattern based on ${{c_{\tilde \kappa}^{(m)}}}$, which consists of periodic uncorrelated sequences used for the TH of pulses. $T_\text c$ is the chip time of the TH sequences, ${c^{(m)}_{\tilde \kappa}} {T_\text c}$ represents the additional time-shift of the $r$-th pulse of the $\kappa$-th data symbol aside from the time-shift ${\xi} {q^{(m)}_\kappa}$ arising from PPM,
$w(t)$ indicates the energy normalized second order Gaussian pulse, i.e.,
\begin{align}
\int_0^{{T_\text b}} {{w^2}\left( t \right)dt} = 1,    \nonumber
\end{align}
where $T_\text b$ is the pulse duration. Considering the first two data symbols, the mapping between the pulse position and data symbols is shown in the axis of {\it TX after TH-PPM} in Fig.~\ref{Signal Propagation Delay}.

The received signals at the agent $n$ can be expressed as \eqref{eqRXAgent}, where $*$ indicates the linear convolution, ${{\alpha}^{(mn)}}$ is the distance attenuation factor, $h(t)$ is the UWB small-scale propagation channel, which could be found in~\cite{UWBChannel} for typical scenarios. The noise at the receiving anchor $z(t)$ is additive white Gaussian noise (AWGN).
To avoid heavy inter-symbol interference, the following relationship of time parameters needs to be satisfied:
$$\xi + {t_{\rm spr}} + t_{\rm TH} < T_{\text f},$$
where ${t_{\rm spr}}$ is the maximum delay spread of the UWB channel, $t_{\rm TH}$ is the maximum time shift length caused by the TH sequence.

\subsection{Asynchronous ICL Method}

Fig.~\ref{Signal Propagation Delay} shows the signal propagation delay from the anchor to the agent. In this paper, we primarily focus on the case depicted in the {\it RX Asynchronous axis} in Fig.~\ref{Signal Propagation Delay}. We define the pseudo-TOA of the $r$-th pulse of the $\kappa$-th data symbol from anchor $m$ to agent $n$ as\footnote {Only the first path in the multipath channel we are interested in this paper.}
\begin{equation} \label{tau}
\begin{aligned}
\tilde t_{\kappa}^{(mn)} \! = & {\Delta \tau^{(mn)} } \! + \! \varepsilon^{(mn)} \big( {\kappa T_{\text f}} \big) \! + \! {\tau^{(mn)}_{\kappa}}
+ \xi {q^{(m)}_\kappa}{+}{c^{(m)}_{\tilde \kappa}} {T_\text c} + {\kappa}{T_{\text f}}, \\
& \:{n \in\mathcal{N}_\text{a}}, {m \in\mathcal{N}_\text{b}}.
\end{aligned}
\end{equation}
The asynchronous ICL problem is to solve ${q^{(m)}_{\kappa}}$ {(\it communication)} and ${\tau^{(mn)}}$ or ${\Delta \tau^{(n)} } + {\tau^{(mn)}_{\kappa}}$
{(\it localization)} in (\ref{tau}) simultaneously.

\subsubsection{Differential sequential detection based data modulation}    \label{SequentialDetection}

The TH sequence, as parameters known to the agent, can be temporarily ignored to simplify the discussion. Then the pseudo-TOA can be written as
\begin{equation}   \label{TOANoTH}
\begin{aligned}
t_{\kappa}^{(mn)} = & {\Delta \tau^{(mn)} } + \varepsilon^{(mn)} \big( {\kappa T_{\text f}} \big) + {\tau^{(mn)}_{\kappa}}
+ \xi {q^{(m)}_\kappa} + {\kappa}{T_{\text f}},  \\
& {n \in\mathcal{N}_\text{a}}, {m \in\mathcal{N}_\text{b}}.
\end{aligned}
\end{equation}

We define $\Delta t_{\kappa+1}^{{(mn)}}$ as the time interval of pseudo-TOA for adjacent data symbols transmitted from anchor $m$ to anchor $n$, expressed as
\begin{equation}
\Delta t_{\kappa+1}^{(mn)} = t_{\kappa + 1}^{(mn)} - t_{\kappa}^{(mn)}.     \label{eq TOAInterval}
\end{equation}

Since PPM modulates data based on pulse position, demodulation of $q^{(m)}_{\kappa+1}$ can be performed according to the interval between adjacent data symbols $\Delta t_{\kappa+1}^{(mn)}$ and the previous data $q^{(m)}_{\kappa}$, which can be expressed as\footnote{The clock drift $\big( \varepsilon^{(mn)}{T_{\text f}} \big)$ between adjacent periods can be neglected.}
\begin{equation}
{q_{\kappa + 1}^{(m)}} = f\Big(\Delta t_{\kappa+1}^{(mn)},{q_{\kappa}^{(m)}}\Big).   \label{demodulation}
\end{equation}
The demodulation function $f(\cdot)$ could be described as a standard {\it sequential estimation} problem, and details will be shown later in~(\ref{sequential_demod}). We define ${q_{\kappa}^{(m)}}$ as the {\it reference symbol} used for the demodulating ${q_{\kappa+1}^{(m)}}$.

{\bf The signal processing in this subsection corresponds to step \textcircled{1} and \textcircled{2} in Fig.~\ref{ICLprocess}.}

\subsubsection{Clock compensation and localization}  \label{Sec Clock}

Clock offset compensation among anchors is prerequisite for agent TDOA localization. The clock offset between anchor 1 and anchor $m$ is given by
\begin{equation}
\Delta \tau_{\text {b}_m} = \Delta \tau^{(1n)} - \Delta \tau^{(mn)},  \quad m = 2,..., N_\text b.    \label{Eq tau alpha}
\end{equation}

The TOA of signals between anchor 1 and anchor $m$ is given by
\begin{equation}
\tau_{\text b_m} = \frac{1}{c}{\left\| {{\mathbf{r}_{1}} - \mathbf{r}_{m}} \right\|_2},  \quad m = 2,..., N_\text b,
\end{equation}
where ${\mathbf{r}_{1}}$ and ${\mathbf{r}_{m}}$ are the positions of anchor 1 and anchor $m$. The local time at which anchor $m$ receives the periodically
signal $w(t)$ transmitted from anchor 1 can be expressed as
\begin{equation}     \label{EqAnchorTime}
t_{\text{b}_m} = \Delta \tau_{\text b_m}+\tau_{\text b_m}.
\end{equation}

The synchronization between anchors can be achieved using \eqref{EqAnchorSyn},
\begin{equation}     \label{EqAnchorSyn}
\Delta \boldsymbol \tau_{\text b} = \arg \min g\big( {\Delta \boldsymbol{\tau}_{\text b}} \big),
\end{equation}
where
$$\Delta \boldsymbol \tau_{\text b} = \big[ \Delta \tau_{\text {b}_2}, ..., \Delta \tau_{\text {b}_{N_\text b}} \big]^{\text T},$$
$$g({\Delta \boldsymbol{\tau}_{\text {b}}}) = \left\| {{\boldsymbol t}_{\text{b}} - \Delta \boldsymbol \tau_{\text b}-{\boldsymbol \tau}_{\text b}}  \right\|,$$
and $\boldsymbol \tau_{\text b} = \big[ \tau_{\text {b}_2}, ...,\tau_{\text {b}_{N_\text b}} \big]^{\text T}$.
Numerical solutions, such as the linear least square (LLS), could be applied to solve \eqref{EqAnchorSyn}. Once the anchors are synchronized, the initial clock offset between all anchors and a specific agent becomes identical. We denote this common clock offset as $\Delta \tau^{(n)}$.

We define the pseudo-delay of the $\kappa$-th symbol between the agent $n$ and anchor $m$ as
\begin{equation}      \label{eq pseudorange}
\begin{aligned}
\beta _{\kappa}^{(mn)} & = {t_{\kappa}^{(mn)}} - {\xi {q_\kappa^{(m)}}}- \kappa T_{\text f}  \\
& \overset {\triangle} = \Delta \tau^{(n)} + {{\tau}^{(mn)}_{\kappa}} + \varepsilon^{(mn)} \big( {\kappa T_{\text f}} \big).
\end{aligned}
\end{equation}

\begin{figure*} [t]
\centerline{\includegraphics[width = 2.0\columnwidth]{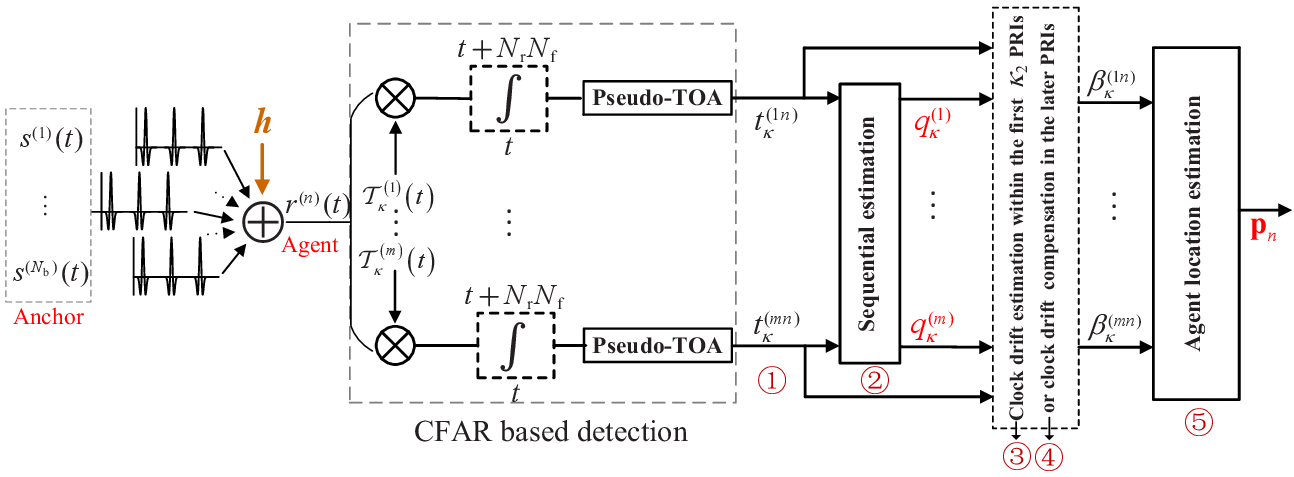}}
\caption{Asynchronous ICL process of agent $n$.}      \label{ICLprocess}
\end{figure*}

\begin{enumerate}
  \item[a)] {\bf Initial agent localization.}
  Assume that no clock drift is present in the initial $\kappa_1$ PRIs between the agent and anchor, i.e., $\varepsilon^{(mn)}\big( {\kappa_1 T_{\text f}} \big)< t_{\text s}$, allowing the clock drift to be neglected.
  Consequently, the pseudo-delay can be expressed as $\dot \beta _{\kappa}^{(mn)} = \Delta \tau^{(n)} + {{\tau}^{(mn)}_{\kappa}}$. Determining the agent's position then reduces to solving \eqref{EqTDOA1}, which is referred to as TDOA-based localization.\footnote{This typically occurs at the initial moment of pulse transmission.}
  \begin{equation}          \label{EqTDOA1}
    c {\mathbf T} = \mathbf{F}(\mathbf p),
  \end{equation}
  where ${\mathbf T} = \Big[\: \dot \beta _{\kappa}^{(2n)}-\dot \beta _{\kappa}^{(1n)}, ..., \dot \beta _{\kappa}^{(4n)}-\dot \beta _{\kappa}^{(1n)} \: \Big]^{\!\text T}$, $\mathbf{F}(\mathbf p) = \Big[\: {\left\| {{\mathbf{r}_2} - \mathbf{p}_n} \right\|_2}-{\left\| {{\mathbf{r}_1} - \mathbf{p}_n} \right\|_2}, ...,{\left\| {{\mathbf{r}_4} - \mathbf{p}_n} \right\|_2}-{\left\| {{\mathbf{r}_1} - \mathbf{p}_n} \right\|_2} \: \Big]^{\!\text T}$.

  {\bf The signal processing in this subsection corresponds to step \textcircled{5} in Fig.~\ref{ICLprocess}.}

  \item[b)] {\bf Clock drift estimation.} Assume that the agent remains stationary during the first $\kappa_2$ PRIs, meaning that $\Delta \tau^{(n)} +
  {{\tau}^{(mn)}_{\kappa}}$ remains constant, and $\varepsilon^{(mn)}\big( {\kappa_2 T_{\text f}} \big) > t_{\text s}$. The clock drift between anchor $m$ and agent $n$ can be estimated by
  \begin{equation}  \label{Eq DriftEst}
  \Big( \beta _{\kappa_2}^{(mn)}-\beta _{\kappa_1}^{(mn)} \Big) = {\varepsilon^{(mn)}}(\kappa_2-\kappa_1){T_{\text f}}.
  \end{equation}

  {\bf The signal processing in this subsection corresponds to step \textcircled{3} in Fig.~\ref{ICLprocess}.}

  \item[c)] {\bf Agent tracking.} Based on the drift estimation results $\xi^{(mn)}$ in {\bf step b)}, agent position tracking can be realized by
  \begin{equation}          \label{EqTDOA}
    c {\mathbf T_{\text {trac}}} = \mathbf{F}(\mathbf p),
  \end{equation}
  where ${\mathbf T_{\text{trac}}} = \Big[\: \beta _{\kappa,{\text t}}^{(2n)}-\beta _{\kappa,{\text t}}^{(1n)}, ..., \beta _{\kappa,{\text t}}^{(4n)}-\beta _{\kappa,{\text t}}^{(1n)} \: \Big]^{\!\text T}$, \\
  $\beta _{\kappa,{\text t}}^{(2n)} = {t_{\kappa}^{(2n)}} - {\xi {q_\kappa^{(2)}}}- \kappa T_{\text f} - \varepsilon^{(2n)} \big( {\kappa T_{\text f}} \big)$.

  {\bf The signal processing in this subsection corresponds to step \textcircled{4} and step \textcircled{5} in Fig.~\ref{ICLprocess}.}

\end{enumerate}

\section{Multi-access Asynchronous Symbol Acquisition and ICL Process}   \label{ICLProcess}

\begin{figure*} [b]
\hrulefill
\begin{equation}                 \label{sequential_demod}
f\Big(\Delta t_{\kappa+1}^{(mn)},{q_{\kappa}^{(m)}} \Big) = \left\{ \begin{array}{l}
 - 1 + {q_{\kappa}^{(m)}},  \hspace{0.1cm}  u_{1,0} - e {\:< \:}\Delta t^{{(mn)}}_{\kappa + 1} \le u_{1,0} + e,
 \:\: \big({q_{\kappa}^{(m)}} = {\text {`} 1 \text{'}} \big) \\    [3pt]
\hspace{0.28cm} 0 + {q_{\kappa}^{(m)}},  \hspace{0.1cm}  u_{0,0} - e {\:< \:}\Delta t^{{(mn)}}_{\kappa + 1} \le u_{0,0} + e  ,
\:\: \big( {q_{\kappa}^{(m)}} = {\text {`} 1 \text{'}} \: {\rm{ or \: {\text {`} 0 \text{'}}}} \big)         \\                    [3pt]
\hspace{0.28cm} 1 + {q_{\kappa}^{(m)}},  \hspace{0.1cm}  u_{0,1} - e {\:< \:}\Delta t^{{(mn)}}_{\kappa + 1} \le u_{0,1} + e  ,
\:\: \big( {q_{\kappa}^{(m)}} = {\text {`} 0 \text{'}} \big)                                                  \\
\end{array} \right.
\end{equation}
\end{figure*}

\subsection{Symbol Detection Based on CFAR}    \label{sec_cfar}

In the ICL UWB system proposed in Sec.~\ref{SecModel}, the pseudo-TOA referred to in \eqref{tau} is an important observation parameter for implementing the ICL function. Accordingly, a matched-filtering (MF) based pulse acquisition solution is provided in this subsection.

The local template pulse corresponding to the TH sequence in the receiver, which is used for detecting pulses transmitted from anchor $m$, is given by
\begin{equation}     \label{eq tempalte}
{\cal T}^{(m)}_{\kappa}(t) = \sqrt{{E_{\text {tb}}}} {\sum\limits_{r = 0}^{N_\text r - 1}  w\Big( t-{r T_{\text f}}-{c^{(m)}_{\tilde \kappa}} {T_\text c} \Big)},
\end{equation}
where $\tilde \kappa= {\kappa{N_\text r}+r}$.

\begin{proposition}   \label{ProCFAR}

The decision variable at time $T$ after the MF of the received signals is expressed as
\begin{equation}       \label{eqrMF}
\begin{split}
{\gamma_{\text {MF}}^{(mn)}}({T})
=\;& \int_{0}^{{N_\text{r}}{T_{\text f}}} {{R^{(m)}}(t){\cal T}^{(n)}_{\kappa}(t)}dt,
\end{split}
\end{equation}
where $R^{(m)}(t) = r^{(m)}(t-T)$.

The threshold for data symbol detection at the agent is given by
\begin{equation}       \label{Eq thred}
\gamma = {Q^{-1}}\big( {{P_{\text {FA}}}} \big) \sqrt {\int_0^{{N_\text{r}}{T_{\text f}}}{E_\text{tb}}{\Big({\cal T}^{(n)}_{\kappa}(t) \Big)^2}{\sigma_z^2}dt} \: ,
\end{equation}
where $Q(\cdot)$ is the right tail function of the standard normal distribution, $P_{\text {FA}}$ is the false alarm probability of pulse detection, $\sigma_z^2$ is the variance of the noise.
\end{proposition}

The general pulse detection probability is
\begin{equation}
\small
{P_\text{D}}\! = \!Q \! \Bigg( \!\! {Q^{{\rm{ - }}1}} \! \big( {{P_\text{FA}}} \big)- {{\alpha^{(mn)}}\int_{0}^{{N_\text{r}}{T_{\text f}}}
{\big({{s}}(t)* h(t)\big){\cal T}^{(n)}_{\kappa}(t)}dt \over {\sqrt{ {\int_{0}^{{N_\text{r}}{T_{\text f}}} {\Big({\cal T}^{(n)}_{\kappa}(t)\Big)^2}{\sigma _z^2}dt} + {N_{\text r}}{\sigma_{\text {a}}^2}\sum\limits_{n = 2}^{N_{\text a}}A^{(mn)}}}} \! \Bigg)\!,        \label{eq PD}
\end{equation}
where ${\sigma _{\text {a}}^2} = {1 \over T_{\text f}}{\int_{0}^{T_{\text f}}\Big[ {\int_{0}^{T_{\text f}}}w(t-\tau)w(t)dt \Big]^2 d\tau}$.

\vspace{3pt}
\begin{Proof}
See Appendix \ref{AppendixProCFAR}.
\end{Proof}

\begin{remark}
The threshold $\gamma$ can be determined based on a preset false alarm probability ${P_\text{FA}}$. If the decision variable satisfies $\gamma_\text{MF}^{(mn)} > \gamma$, it indicates the presence of a data symbol. For instance, as illustrated by the {\it TH template signal ${\cal T}^{(m)}_{0}$} in Fig.~\ref{Signal Propagation Delay}, ${\cal T}^{(m)}_{0}$ serves as the template signal for detecting the first symbol transmitted by the anchor $m$. During the matched filtering (MF) process, when the template pulse aligns with the received signal at the position $t_0^{(mn)}$ shown in Fig.~\ref{Signal Propagation Delay}, such that $\gamma_\text{MF}^{(mn)} > \gamma$, the first data symbol is considered to have been successfully detected.
\end{remark}

\begin{figure*} [t]
\centering
\centerline{\includegraphics[width = 1.8\columnwidth]{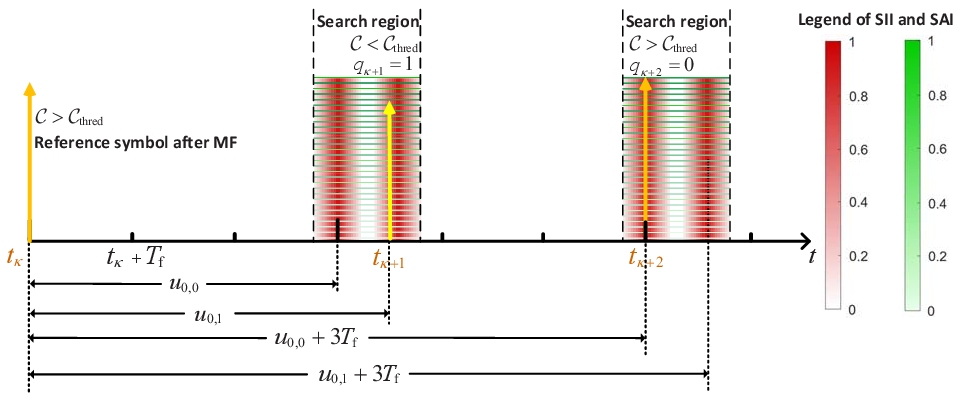}}
\caption{Symbol detection based on confidence metric $\big(N_{\text r} = 3. \: \text {The superscript}  \text {is omitted for simplify} \big)$.}
\label{Fig SoftInformation}
\vspace{-10pt}
\end{figure*}

\subsection{ICL Implementation}    \label{SecICL}

\begin{figure} [h]
\centerline{\includegraphics[width = 0.9\columnwidth]{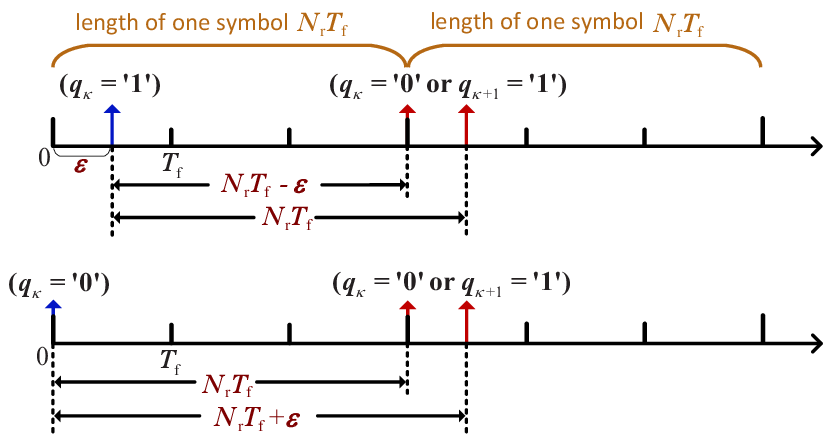}}
\caption{Position of adjacent data symbol after MF. (The superscript of $q_{\kappa}^{(m)}$ has been omitted for simplicity.)}      \label{Fig AdjentInterval}
\end{figure}

Without loss of generality, the data symbols carried using PPM can be characterized into following four situations: $\left\{ {10} \right\},\left\{ {11} \right\},\left\{ {00} \right\},\left\{ {01} \right\}$. A simplified diagram for all four situations is given in Fig.~\ref{Fig AdjentInterval}.\footnote{The pseudo-TOA obtained based on the matched filtering (MF) process does not include the time shift of the TH sequence, consistent with the pseudo-TOA expression given in \eqref{TOANoTH}.}

The expectations of the interval of adjacent data symbols \big($\Delta t_{\kappa+1}^{(mn)}$ \big) are given as follows:
\begin{equation}     \label{Eq SII expectation}
  \begin{split}
   &  u_{0,0} \overset {\triangle} = {u_{1,1}} ={N_\text r T_{\text f}},  \:\quad   \\
   &  u_{0,1} = {N_\text r T_{\text f}} + \xi,     \:\quad
      u_{1,0} = {N_\text r T_{\text f}} - \xi,
   \end{split}
\end{equation}
where $u_{0,0}$ denotes the expectation of the interval in the case where $q_{\kappa}^{(m)} = \text{`0'} $ and $q_{\kappa+1}^{(m)} = \text{`1'}$.

We set the fault-tolerance space for the position of the data symbol detection as $e = {1\over 2}\xi$.
As mentioned in (\ref{demodulation}), we could define the demodulation function $f\Big( \Delta t_{\kappa+1}^{(mn)},{q_{\kappa}^{(m)}} \Big)$ as (\ref{sequential_demod}) according to the various cases in ~(\ref{Eq SII expectation}).

After data demodulation and clock drift compensation based on Sec.~\ref{Sec Clock}, $\beta_{\kappa}^{(mn)}$ can be obtained according to~(\ref{eq pseudorange}). Subsequently, the position of the agent is estimated based on \eqref{EqTDOA}. In this paper, the root-mean-square error (RMSE) of the agent is employed to evaluate the positioning performance.
\begin{equation}
{\rm{RMSE  = \Big(}}\mathbb{E}\left\| {{\bf{p}}_{n} - {{{\bf{\hat p}}}_{n}}} \right\| \Big)^{{1 \mathord{\left/{\vphantom {1 2}} \right.
\kern-\nulldelimiterspace} 2}},
\end{equation}
where ${{\bf{\hat p}}}_{n}$ represents the estimated position of the agent $n$. The overall implementation structure of the ICL system at the agent $n$ is illustrated in Fig.~\ref{ICLprocess}.

\section{Symbol Detection Enhancement based on Confidence Metric}     \label{sec_soft}

\begin{figure*}[b]
\hrulefill
\begin{equation}   \label{EqPD2}
\begin{aligned}
{P_{{\text{D2}}}} = \; &{C_{N_{\text{sf}_2}}^{N_{\text{suc}}}{{\big(P_{\text{D}}\big)}^{N_{\text{suc}}}}
{{\big(1 - P_{{\text{D}}}\big)}^{({N_{\text{sf}_2}} - {N_{\text{suc}}})}}}
+ C_{N_{\text{sf}_2}}^{N_{\text{suc}}+1} {\big(P_{\text{D}}\big)^{({N_{\text{suc}}} + 1)}}
{\big(1 -P_{\text{D}}\big)^{({N_{\text{sf}_2}} - {N_{\text{suc}}}-1)}}   +   \cdots   +
& C_{N_{\text{sf}_2}}^{N_{\text{sf}_2}}{\big(P_{\text{D}}\big)^{N_{\text{sf}_2}}}{\big(1 - P_{\text{D}}\big)^0}
\end{aligned}
\end{equation}
\end{figure*}

Accurate detection of data symbols is essential in the sequence detection process outlined in \eqref{sequential_demod}. However, pulse false alarms or missed detections can easily cause symbol misalignment. For example, if the agent fails to detect the $\kappa$-th data symbol, it may incorrectly interpret the $(\kappa+1)$-th data symbol as the $\kappa$-th. {\bf Such misalignment can significantly degrade performance in communication, clock drift compensation, and localization.}

\subsection{Frame Structure and SFD Design}

In the sequential detection-based demodulation process, the SFD, particularly for the first symbol, functions as a reference symbol and serves as an overhead to eliminate the need for synchronization at the receiver. This subsection examines the SFD segment in detail.
\begin{figure} [h]
\centerline{\includegraphics[width = 1\columnwidth]{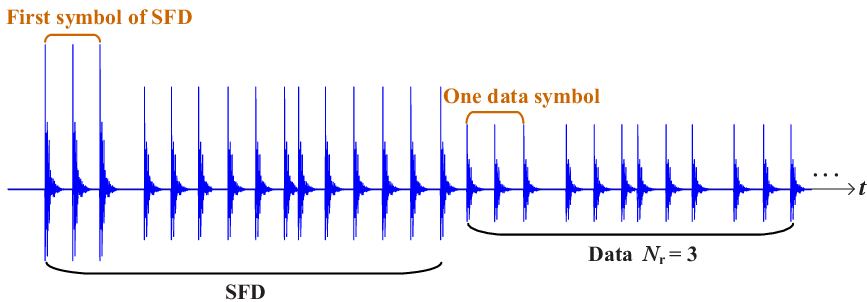}}
\caption{The proposed frame structure.}    \label{Fig Frame Structure}
\end{figure}

As shown in Fig.~\ref{Fig Frame Structure}, the frame structure is divided into three segments: the first segment consists of the initial symbol, the second segment comprises the SFD excluding the initial symbol, and the third segment contains only the data symbols.
The prerequisite for identifying the SFD is the successful detection of the first symbol, whereas the second segment of the SFD tolerates a maximum symbol error rate $P_{\text e}$. The detection probability of the second segment is given by \eqref{EqPD2}, where $N_{\text {sf}_2}$ denotes the number of symbols of the second segment of the SFD, ${N_{{\text{suc}}}} = \left\lfloor {N_{\text {sf}_2}}(1-{P_{\text e}})\right\rfloor$ represents the minimum number of symbols required to be detectable in the second segment of SFD, $C_{N_{{\text {sf}}_2}}^{N_\text {suc}}$ represents the combination function, specifically the number of ways to choose $N_{\text {suc}}$ elements from a total of $N_{\text {sf}_2}$ elements.

Consequently, the overall detection probability of the SFD is given by the product of the detection probabilities of the first symbol and the second segment, expressed as
\begin{equation}    \label{EqSFDDtec}
P_{\text {FD}} = P_{\text D}P_{\text {D2}}.
\end{equation}

We denote the pseudo-TOA of SFD from anchor $m$ at the agent $n$ as $t^{(mn)}_{-1}$.

\subsection{Confidence Based Differential Data Demodulation}     \label{sec head enhancement}

In this subsection, we address the issue of symbol misalignment by introducing a symbol acquisition enhancement method based on {\it soft information (SI)} during symbol propagations.
We further decompose the likelihood-based SI into two distinct parts: the soft interval information (SII), which pertains to the interval
$\Delta t^{(mn)}_{\kappa + 1}$ between consecutive symbols, and the soft amplitude information (SAI), which is connected to the decision variable ${\gamma_{\text{MF}}^{(mn)}}$ of the MF.

The greater the extent to which the MF output exceeds the threshold $\gamma$, the higher the confidence in symbol detection, and vice versa. Accordingly, the SAI can be defined as
\begin{equation}           \label{Eq_SAI}
\begin{split}
{\cal S}\left( {{\gamma _{{\text{MF}}}}} \right) = \left\{ \begin{array}{l}
{{{\gamma _{{{{\scalebox{0.5} {MF}}}}}}} \over \gamma },   \quad\quad  {{\gamma _{{\rm{\scalebox{0.5}{MF}}}}} < \gamma},   \\
 1,   \quad\quad\quad    {\gamma _{{\rm{\scalebox{0.5}{MF}}}}} \ge \gamma.
\end{array} \right.
\end{split}
\end{equation}

It is reasonable to consider the soft confidence function of $\Delta t^{{(mn)}}_{\kappa + 1}$  as being consistent with the TOA errors, i.e., a Gaussian function, and we define it as\footnote{The superscript $(mn)$ is omitted for simplicity.}
\begin{equation}
\mathcal{Q}\Big(\Delta t_{\kappa + 1}|q_{\kappa},q_{\kappa+1}\Big) \sim N\Big( {{u_{{q_{\kappa}},{q_{\kappa+1}}}},
\sigma^{\scalebox{0.5}2}{\kern-0.5em}_{\scalebox{0.5}{$\Delta t$}}  } \Big). \label{Q_interval}
\end{equation}

It is assumed that the probabilities of transmitting `0' or `1' in ${q_{\kappa+1}}$ are equal, each being $1 \over 2$.
The normalized confidence of the SII, denoted as ${\mathop{\mathcal L}\nolimits} \left( {\Delta t_{\kappa + 1}} \right)$, is written as follows:
\begin{equation}            \label{eqSII}
\begin{split}
{\mathop{\mathcal L}\nolimits} \Big( {\Delta t_{\kappa + 1}} \Big)
\! = & \exp \! \Bigg( \! { - {{{{\big( {{\Delta t_{\kappa + 1}} - {u_{{q_{\kappa}},0}}} \big)}^2}}
 \over {2 \sigma^{\scalebox{0.5}2}{\kern-0.5em}_{\scalebox{0.5}{$\Delta t$}}}}}\! \Bigg)  \\
 & \! + \exp \! \Bigg( \!  { - {{{{\big( {{\Delta t_{\kappa + 1}} - {u_{{q_{{\kappa}},1}}}} \big)}^2}} \over {2\sigma^{\scalebox{0.5}2}{\kern-0.5em}_{\scalebox{0.5}{$\Delta t$}}}}} \!   \Bigg),  \!\!
\end{split}
\end{equation}
where ${\sigma^{\scalebox{0.5}2}{\kern-0.5em}_{\scalebox{0.5}{$\Delta t$}}} = {{{2c^2}} \over {8{\pi ^2}B^2{SNR}}}$, $B$ is the signal transmission bandwidth, $SNR$ is the ration of the signal power to the noise power.\footnote{The variance is determined with consideration of TOA estimation.}

The normalized confidence of SI, considering both SAI and SII, is expressed as
\begin{equation}  \label{Eq_SI}
\mathcal{C} = \mathcal{S}\mathcal{L}.
\end{equation}

\subsubsection{Symbol decision with confidence metric}

Fig.~\ref{Fig SoftInformation} shows the symbol detection process based on confidence metric. By integrating SII and SAI, the reliability of the current symbol is evaluated. We denote the threshold of the high confidence area as ${\mathcal{C}}_{\text{thres}}$. Specifically, if the confidence $\mathcal{C}$ of the $\kappa$-th symbol satisfies $\mathcal{C}>{\mathcal{C}}_{\text{thres}}$, the $\kappa$-th symbol can be regarded as the reference symbol for the subsequent data demodulation.

\begin{itemize}
\item Use the symbol at $t_\kappa$ as the reference symbol and search for the ${(\kappa+1)}$-th symbol within the designated search region. If the maximum confidence value of $\mathcal{C}$ within the search region satisfies $\mathcal{C}_{\text {max}}> \mathcal{C}_\text {thres}$, the position corresponding to $\mathcal{C}_{\text {max}}$ namely $t_{\kappa+1}$. In this case, the ${(\kappa+1)}$-th symbol cannot be considered as the reference symbol for detecting the $(\kappa+2)$-th symbol.
\item If the confidence value of the $\kappa+1$-th symbol satisfies $\mathcal{C}<\mathcal{C}_\text {thres}$, the $\kappa$-th symbol at $t_\kappa$ is used as the reference symbol to search for the ${(\kappa+2)}$-th symbol within the designated search region. $q_{\kappa+2}$ can be demodulated based on
    \eqref{Eq InterTOA2} and \eqref{sequential_demod}.
    \begin{equation}
     \Delta t_{\kappa + 1} = t_{\kappa+1} - t_{\kappa-1}. \label{Eq InterTOA2}
    \end{equation}
    The expectation in (\ref{eqSII}) is increased by $N_{\text r}{T_{\text f}}$, ensuring that $u_{0,0} = {u_{1,1}} = 2{N_{\text r}{T_{\text f}}}$.
\end{itemize}

The data demodulation process based on confidence metric is presented as {\bf Algorithm~\ref{ICLAlgo}}.
\begin{algorithm} [h]
	\caption{Confidence based differential demodulation process.}   \label{ICLAlgo}
	\begin{algorithmic}[1]
        \STATE Global clock synchronization between anchors based on \eqref{Eq tau alpha}-\eqref{EqAnchorSyn}.
        \STATE SFD $\Big( t^{(mn)}_{-1} \Big)$ acquisition.
        \FOR{$\kappa = 0$ to $N_{\text f}-1$}
            \STATE Estimate the pseudo-TOA of the ${\kappa}$-{th} data symbol $t^{(mn)}_{\kappa}$ in the $\kappa$-th search region.
            \STATE  Get the confidence of the $\kappa$-th data symbol based on \eqref{Eq_SI}.
            \STATE Assume that the $\kappa_{\text H}$-th $(\kappa_{\text H}<\kappa)$ data symbol is the preceding data symbol with a high level of confidence
            \big( $\mathcal{C}>\mathcal{C}_\text{thres}$ reference symbol \big ) before the $\kappa$-th symbol. Perform demodulation of $q^{(m)}_{\kappa}$ according to $\Delta t^{(mn)}_{\kappa}$ based on \eqref{sequential_demod}.
            \IF  {$q_{\kappa_{\text H}}^{(m)} = 1$     \: and \: $\Delta T_{\text f} + u_{1,0}-e < \Delta \tau^{(mn)}_{\kappa} < \Delta T_{\text f} + u_{1,0}+e $}
                 \STATE $q_{\kappa}^{(m)}=0$,  where $\Delta T_{\text f} = (\kappa-\kappa_{\text H}-1)N_{\text r}T_{\text f}$
            \ELSIF {$q_{\kappa_{\text H}}^{(m)} = 1$  \: and \: $\Delta T_{\text f} + u_{1,1}-e < \Delta \tau^{(mn)}_{\kappa} < \Delta T_{\text f} + u_{1,1}+e $}
                    \STATE $q_{\kappa}^{(m)}=1$,
            \ELSIF {$q_{\kappa_{\text H}}^{(m)} = 0$  \: and \: $\Delta T_{\text f} + u_{0,0}-e < \Delta \tau^{(mn)}_{\kappa} < \Delta T_{\text f} + u_{0,0}+e $}
                    \STATE $q_{\kappa}^{(m)}=0$,
            \ELSIF {$q_{\kappa_{\text H}}^{(m)} = 0$  \: and \: $\Delta T_{\text f} + u_{0,1}-e < \Delta \tau^{(mn)}_{\kappa} < \Delta T_{\text f} + u_{0,1}+e $}
                    \STATE $q_{\kappa}^{(m)}=1$,
            \ENDIF
        \ENDFOR
	\end{algorithmic}
\end{algorithm}

\subsubsection{Confidence-based MWLS for clock drift estimation}

The estimation of clock drift can be performed based on \eqref{Eq DriftEst} with the least squares (LS) method as
\begin{equation}      \label{LSDrift}
{\varepsilon ^{(mn)}} = {\Big( {{{\bf{X}}^{\rm{T}}}{\bf{X}}} \Big)^{\scriptsize{-1}}}{{\bf{X}}^{\rm{T}}}{\bf y},
\end{equation}
where ${\bf y} = {\left[ {{\beta _{{\kappa _{\tilde N}}}^{(mn)} - \beta _{{\kappa _{\tilde N - 1}}}^{(mn)},} \:
{\cdots,}  \: {\beta _{{\kappa _2}}^{(mn)} - \beta _{{\kappa _1}}^{(mn)}}  } \right]^{\rm{T}}}$,
${\bf X} = {\Big[
{\big({\kappa _{\tilde N}} - {\kappa _{\tilde N - 1}}\big){T_{\rm{f}}},} \: { \cdots ,} \:{\big({\kappa _2} - {\kappa _1} \big){T_{\rm{f}}}} \Big]^{\rm{T}}}$,
${\kappa_{\tilde N}}$ indicates the number of PRIs utilized for clock drift estimation.

Moreover, leveraging the proposed symbol detection strategy based on SI, the confidence levels of the detected symbols can serve as weights, enhancing the LS approach to form MWLS. Specifically, if the confidence level of a symbol $\mathcal{C}$ is less than the threshold $\mathcal{C}_{\text{thres}}$, the corresponding WLS weight is set to zero; otherwise, $\mathcal{C}$ is used as the weight. Details are presented as
\begin{equation}      \label{WLSDrift}
{\varepsilon ^{(mn)}} = \Big( {{{\bf{X}}^{\rm{T}}}{\bf{W}}{\bf{X}}} \Big)^{\scriptsize{-1}}{{\bf{X}}^{\rm{T}}}{\bf{W}}{\bf y},
\end{equation}
where the weighted vector is expressed by ${\bf{W}} = \Big[{W_{\kappa_{\tilde N}}}, ..., {W_{\kappa_{\tilde n}}}, ...,  {W_{\kappa_{1}}} \Big]^{\text T}$,
\begin{equation}           \label{Eq_SAI}
\begin{split}
{W_{\kappa_{\tilde n}}} = \left\{ \begin{array}{l}
\mathcal{C},   \quad  {\mathcal{C} > \mathcal{C}_\text{thres}} \: \text{of the} \: {\kappa_{\tilde n}}\text{-th symbol},   \\
 0,   \quad    {\mathcal{C} < \mathcal{C}_\text{thres}} \: \text{of the} \: {\kappa_{\tilde n}}\text{-th symbol}.
\end{array} \right.
\end{split}
\end{equation}

\section{Simulations and Experimental Results}    \label{section Numerical}

In this section, simulation results are presented to show the SFD detection performance based on the proposed SFD structure. The effectiveness of the proposed SI algorithm in the multi-access network is evaluated based on clock drift estimation, communication, and localization performance. Gold sequences are selected as the TH sequence \cite{Gold}. Parameters setting are listed as Table.~\ref{parameters}.\footnote{The unit of clock drift $\emph{ppm}$ means `Parts Per Million'.}
\begin{table*} [t]
 \centering
 \setlength{\belowcaptionskip}{7pt}
 \renewcommand{\arraystretch}{1}
 \caption{Simulation parameters}   \label{parameters}
 \begin{tabular}{lll}
  \toprule
  Parameters            & Description                                 & Value    \\
  \midrule
  $f_{\text s}$         & Sampling frequency (\text GHz)              & $10$     \\
  $T_{\text f}$         & PRI (\text ns)                              & $160$    \\
  $\xi$                 & Time-shift of PPM modulation (\text ns)     & $45$     \\
  $P_{\text {FA}}$      & False alarm probability     & $10^{-3}$     \\
  $\mathbf{r_1}$             & Anchor 1 position      & $(2,5,2)$     \\
  $\mathbf{r_2}$             & Anchor 2 position      & $(4,8,3)$     \\
  $\mathbf{r_3}$             & Anchor 3 position      & $(5,5,3)$     \\
  $\mathbf{r_4}$             & Anchor 4 position      & $(7,3,2)$     \\
  $\xi^{(11)}-\xi^{(41)}$    & Clock drift between the four anchors to the agent $\emph{ppm}$  & $(20, 10, 30, 20)$     \\
  \bottomrule
 \end{tabular}
\end{table*}

\subsection{Performance Analysis of SFD}

\begin{figure} [t]
\centerline{\includegraphics[width = 1\columnwidth]{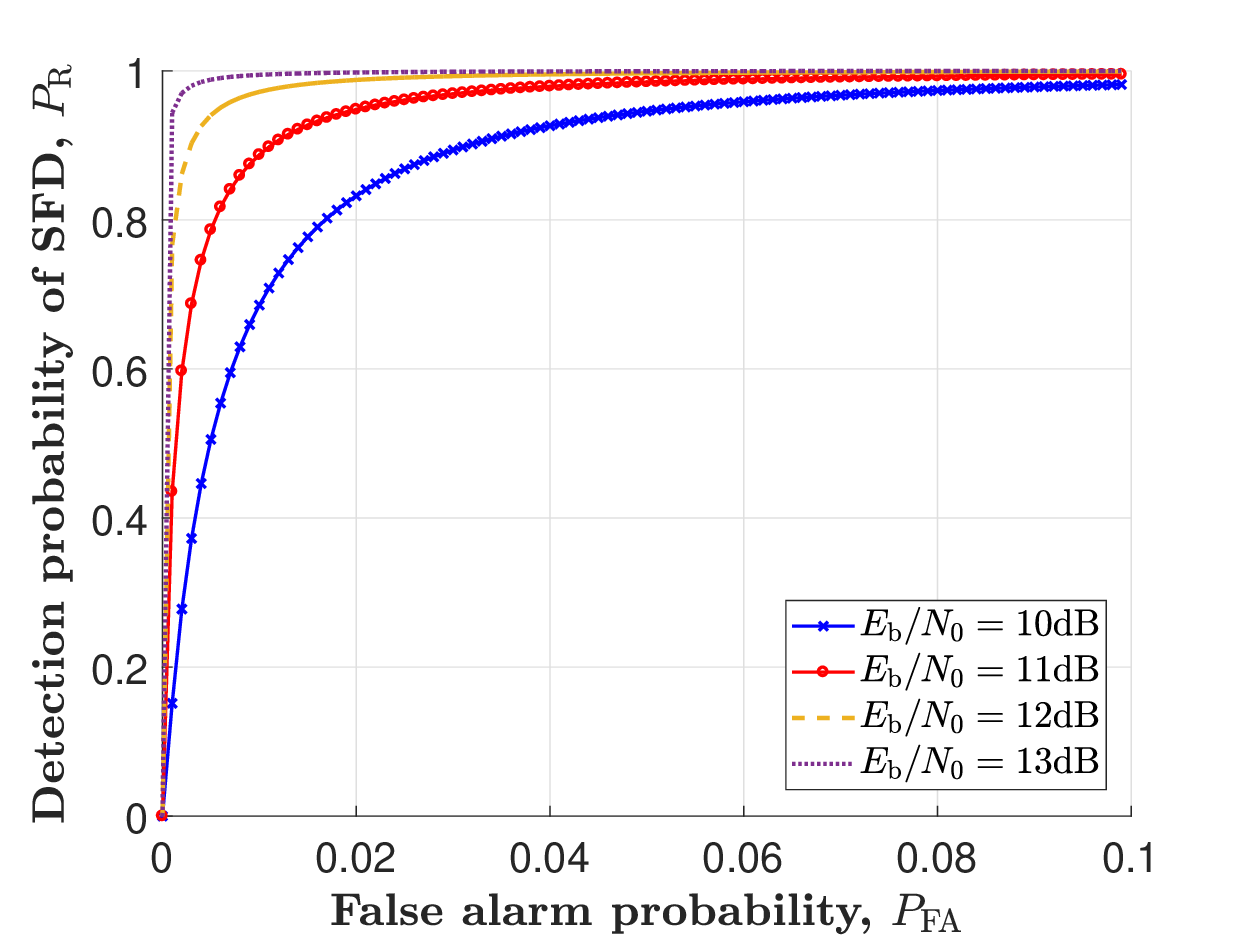}}
\caption{ROC of pulse detection \big($N_{\text r} = 3$\big).}
\label{ROCSNR}
\end{figure}

\begin{figure} [t]
\centerline{\includegraphics[width = 1\columnwidth]{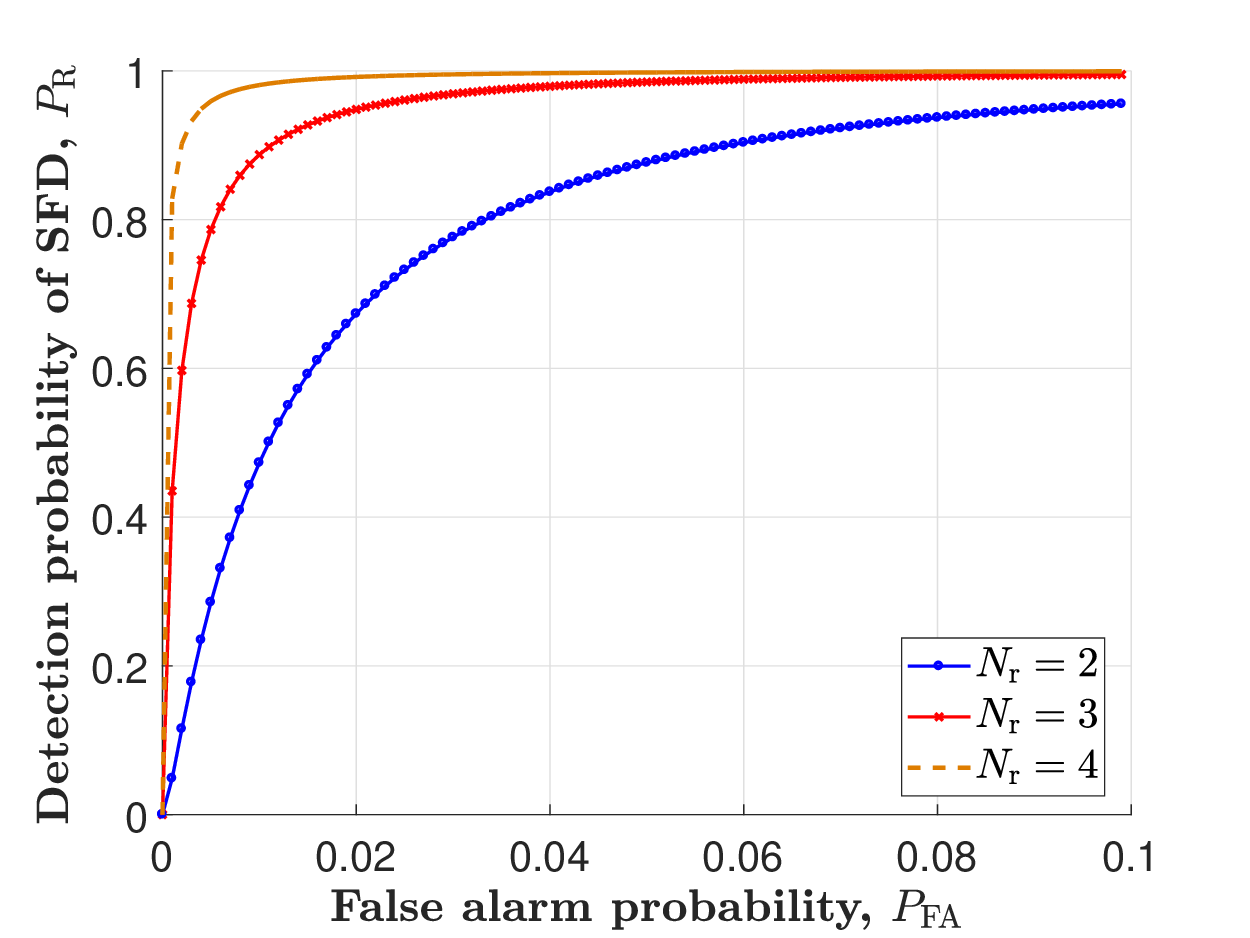}}
\caption{ROC of pulse detection \big($ E_{\text b}/N_0 = 11 \text {dB} $\big).}
\label{ROCNr}
\end{figure}

The maximum symbol error rate for SFD is set to $P_{\text e} = 0.1$, with the total number of SFD symbols defined as $N_{\text {sf}_2}+1 = 8$.
Fig.~\ref{ROCSNR} and \ref{ROCNr} illustrate the relationship between the SFD detection probability and the symbol false alarm probability, represented by the receiver operating characteristic (ROC) curve.
In the SFD detection process,  we aim to maximize the detection probability while minimizing the false alarm probability. Therefore, the upper left region of the ROC curve is of particular interest. For a given $E_{\text b}/N_0$ and $N_{\text r}$, this region can be achieved by adjusting the parameter $\gamma$.
SFD detection performance improves as $E_{\text b}/N_0$ and $N_{\text r}$ increase.

\begin{figure} [t]
\centerline{\includegraphics[width = 1\columnwidth]{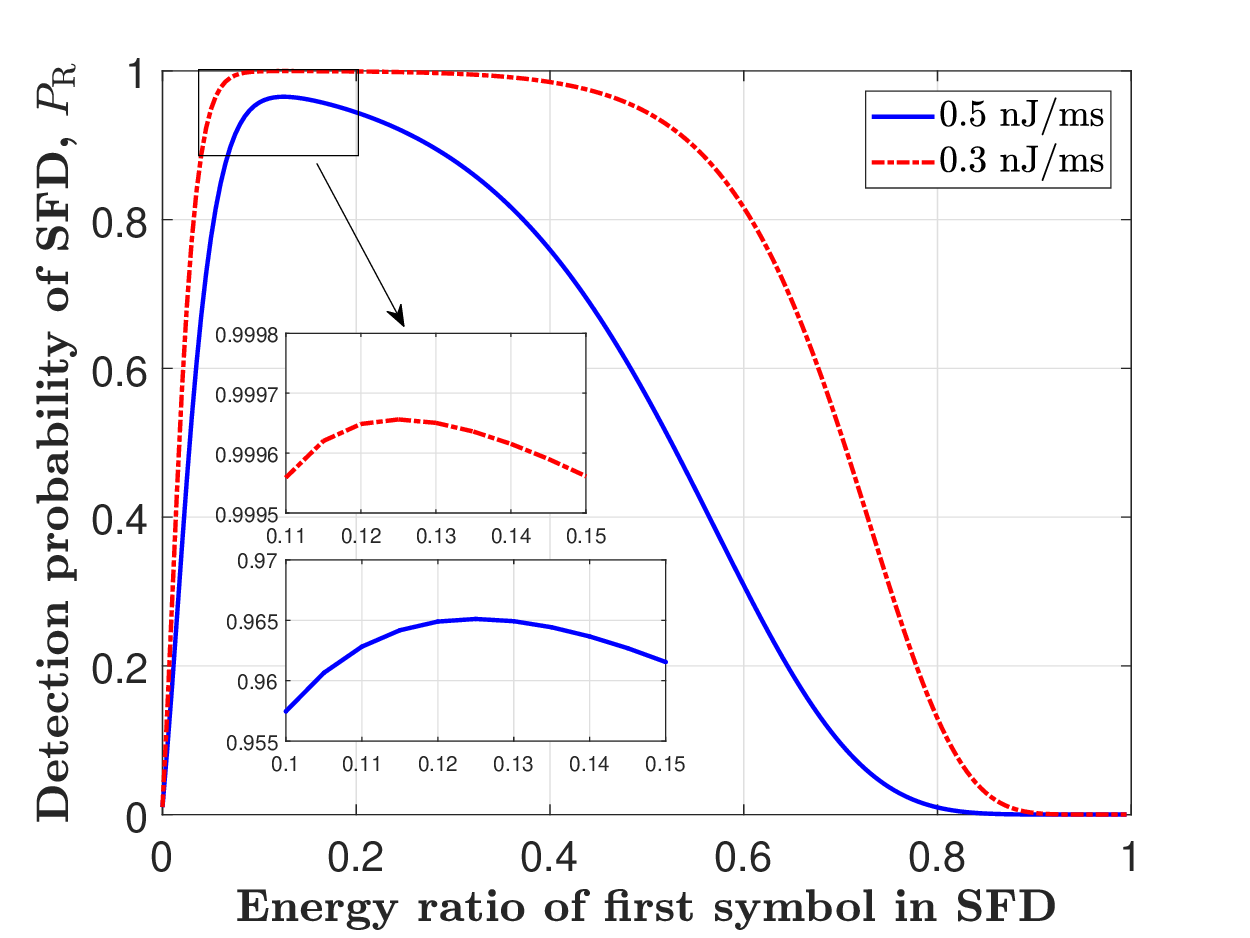}}
\caption{Pulse detection probability ($N_{\text r} = 3$).}
\label{PDSFD}
\end{figure}

Fig.~\ref{PDSFD} shows the SFD detection performance about the distribution of energy between the first symbol and the remaining symbols within the SFD structure. When the total energy of the SFD remains constant, the SFD detection probability reaches its maximum when the first symbol accounts for 12.5\% of the total SFD energy (i.e., when the energy is equally distributed among all symbols in the SFD).

\subsection{Clock Drift, Communication, and Localization with Confidence Metric}

\begin{figure} [t]
\centerline{\includegraphics[width = 1\columnwidth]{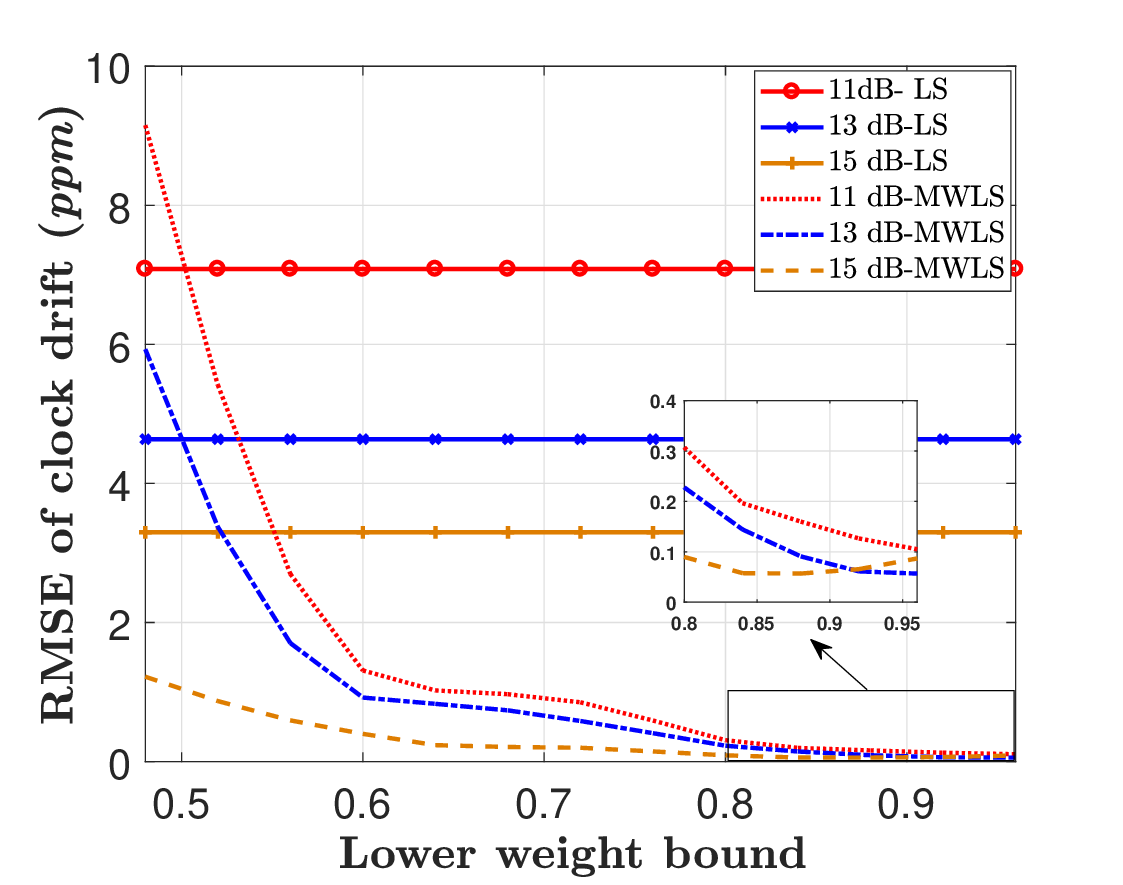}}
\caption{Clock drift estimation ($N_{\text r} = 3$).}
\label{DriftEst}
\end{figure}

Fig.~\ref{DriftEst} shows the RMSE of clock drift estimation considering the confidence threshold $\mathcal{C}_{\text{thres}}$.
The results show that the confidence-based WLS algorithm enhances the accuracy of clock drift estimation.
Moreover, as the confidence threshold $\mathcal{C}_{\text{thres}}$ increases, the estimation performance continues to improve. Notably, when the confidence threshold falls within the range of 0.9 to 0.95, the clock drift estimation stabilizes and reaches a plateau.

\begin{figure} [t]
\centerline{\includegraphics[width = 1\columnwidth]{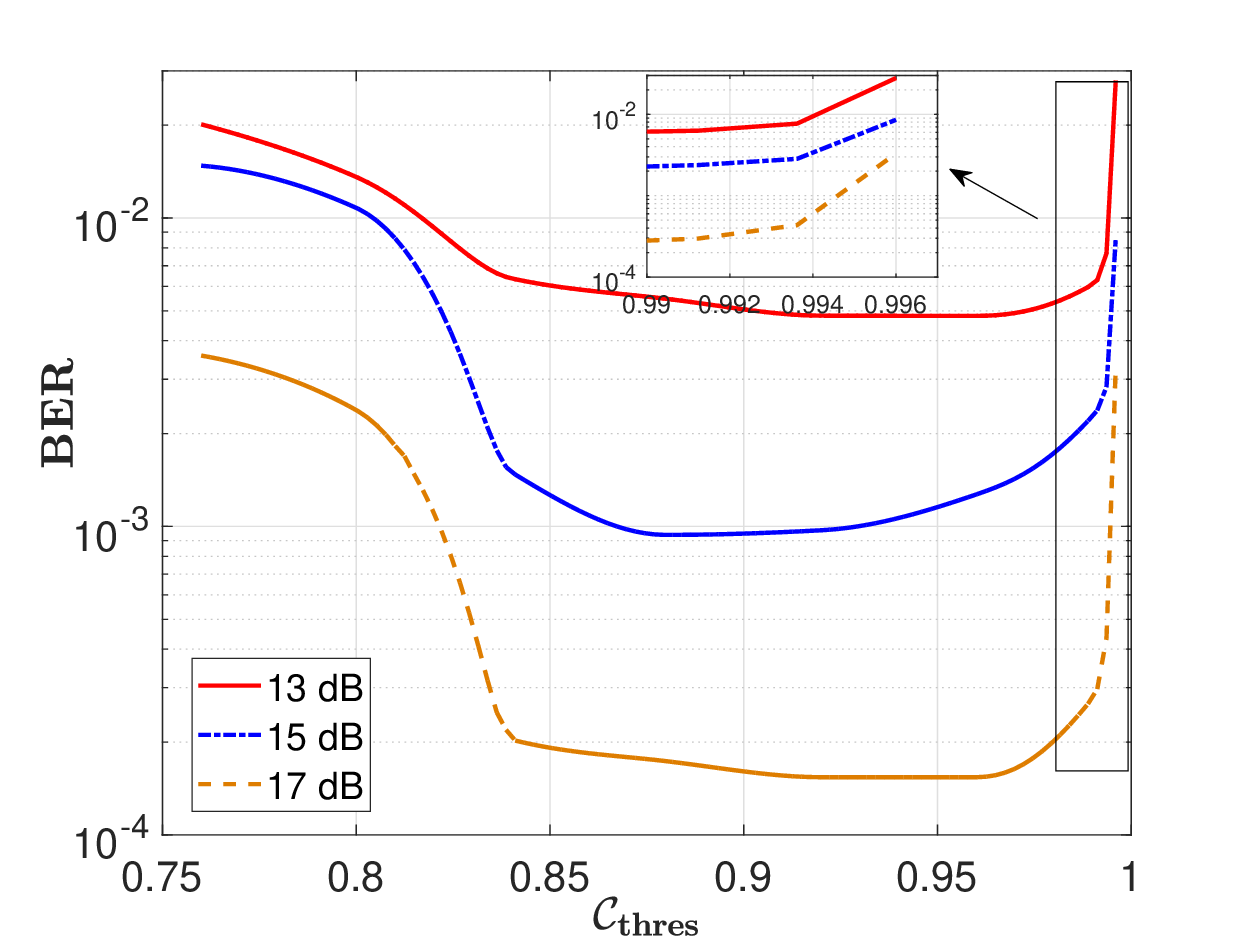}}
\caption{Effect of confidence threshold on communication performance ($N_{\text r} = 3$).}
\label{BERwithConfidence}
\end{figure}

\begin{figure} [t]
\centerline{\includegraphics[width = 1\columnwidth]{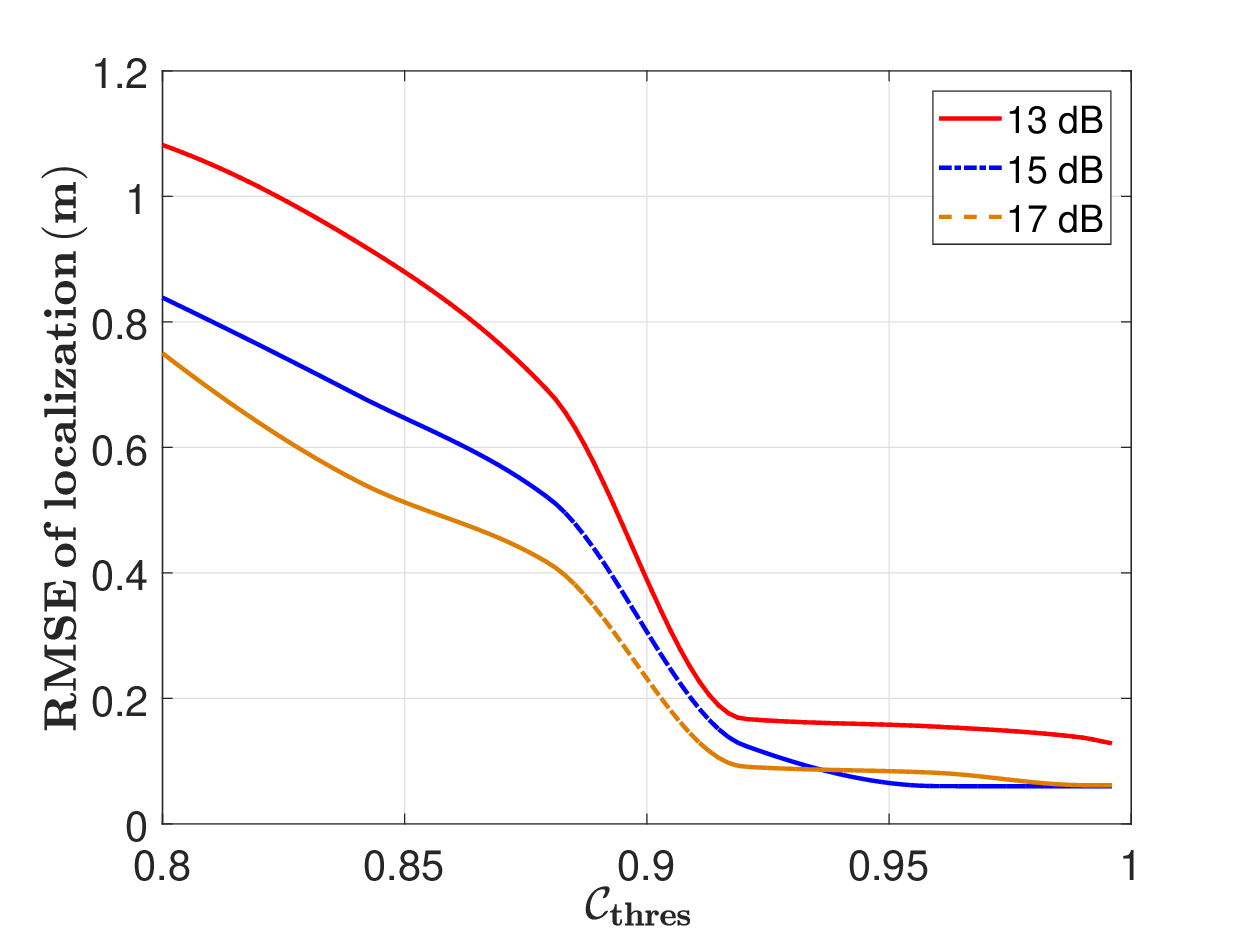}}
\caption{Effect of confidence threshold on localization performance ($N_{\text r} = 3$).}
\label{RMSEwithConfidence}
\end{figure}

Fig.~\ref{BERwithConfidence} and~\ref{RMSEwithConfidence}  illustrate the communication and localization performance as functions of the confidence threshold, with performance measured by the bit error rate (BER) for symbol transmission and the RMSE for localization, respectively. When the confidence threshold is between approximately 0.85 and 0.96, the system achieves a low bit error rate. For confidence thresholds above 0.92, localization performance stabilizes, and the RMSE of localization remains below 0.1m when $E_{\text b}/N_{\text 0}>15 \: \text{dB}$.
However, when $\mathcal{C}_{\text {thres}}>0.993$, the BER starts to increase. This is due to the increasing sparsity of {\it reference symbols} as $\mathcal{C}_{\text {thres}}>0.993$ rises. In the differential sequential detection process, clock drift caused by symbols that are temporally distant can affect the confidence determination of the symbols, leading to a degradation in communication performance.

\begin{figure} [t]
\centerline{\includegraphics[width = 1\columnwidth]{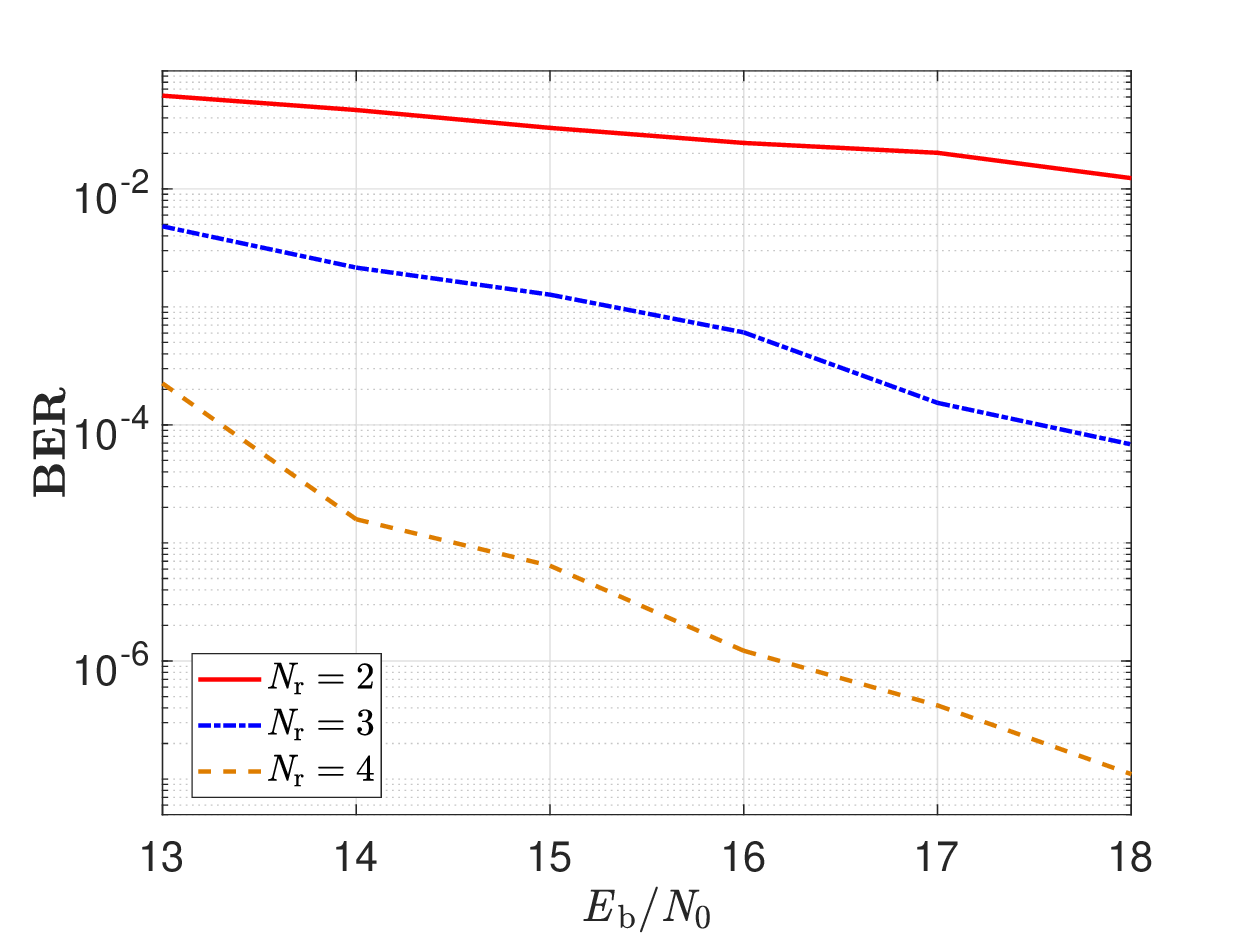}}
\caption{Communication performance under different bit repetition count \big($\mathcal{C}_{\text{thres}} = 0.96$\big).}
\label{BERNr}
\end{figure}

\begin{figure} [t]
\centerline{\includegraphics[width = 1\columnwidth]{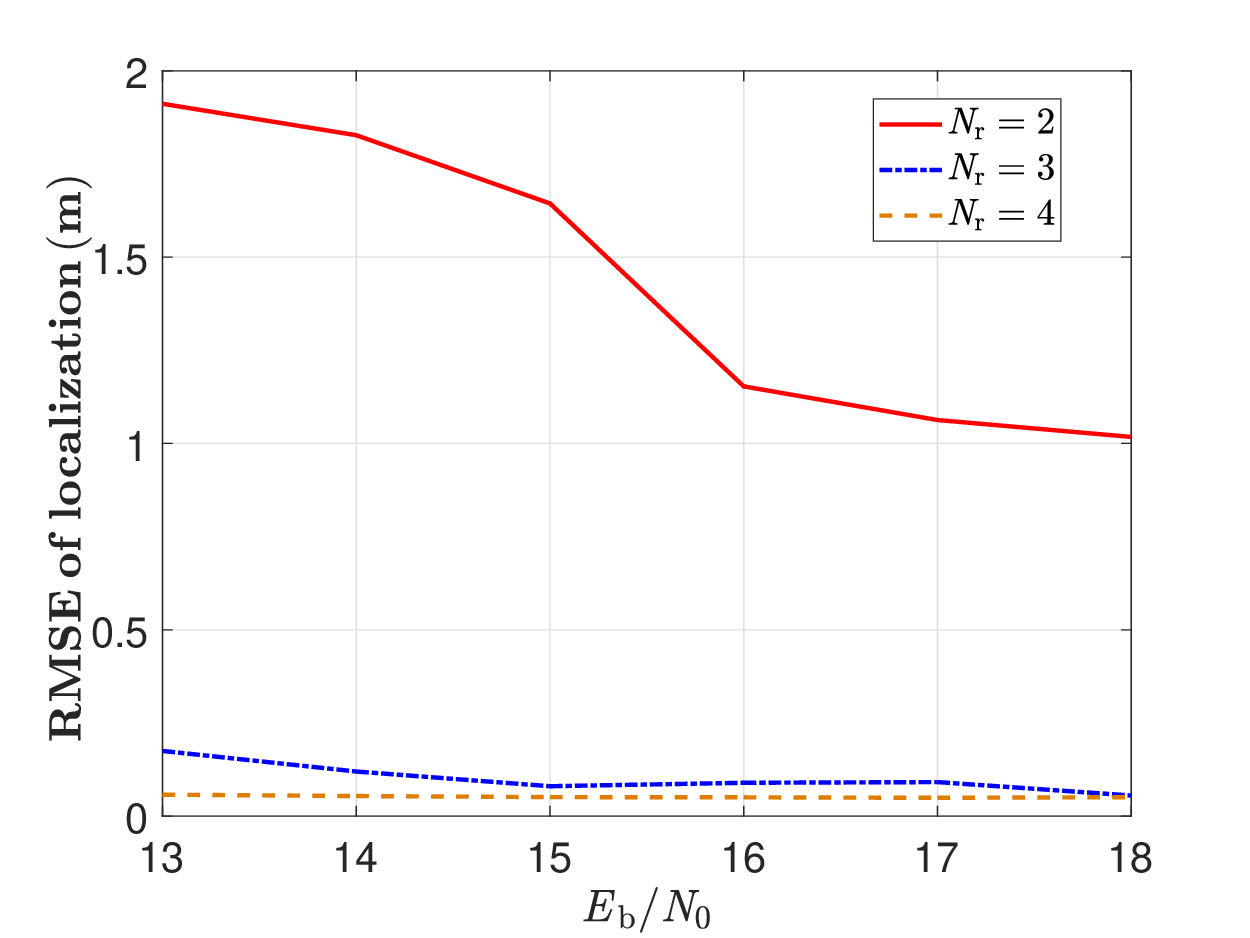}}
\caption{Localization performance under different bit repetition count \big($\mathcal{C}_{\text{thres}} = 0.96$\big).}
\label{MSENr}
\end{figure}

The communication and localization performance as functions of the bit repetition count $N_{\text r}$ are presented in Fig.~\ref{BERNr} and~\ref{MSENr}.
Observed that, as $N_{\text r}$ increases, both the communication BER and localization RMSE decrease.
When the energy of each symbol is fixed, doubling the energy of a single symbol is equivalent to either doubling $N_{\text r}$ or increasing the $E_{\text b}/N_{\text 0}$ by 3 dB. As illustrated in Fig.~\ref{BERNr}, an increase in $N_{\text r}$  has a more significant effect on reducing the BER.
This is because increasing the $E_{\text b}/N_{\text 0}$ does not mitigate the non-orthogonal interference among signals from multiple anchors.
The localization performance shown in Fig.~\ref{MSENr} follows the same pattern.

\subsection{Experiment Measurement}       \label{section Experiment}

\begin{figure*} [t]
\centering
\subfigure[Experimental environment]{\label{FigIndoorExp}\includegraphics[width = 1\columnwidth] {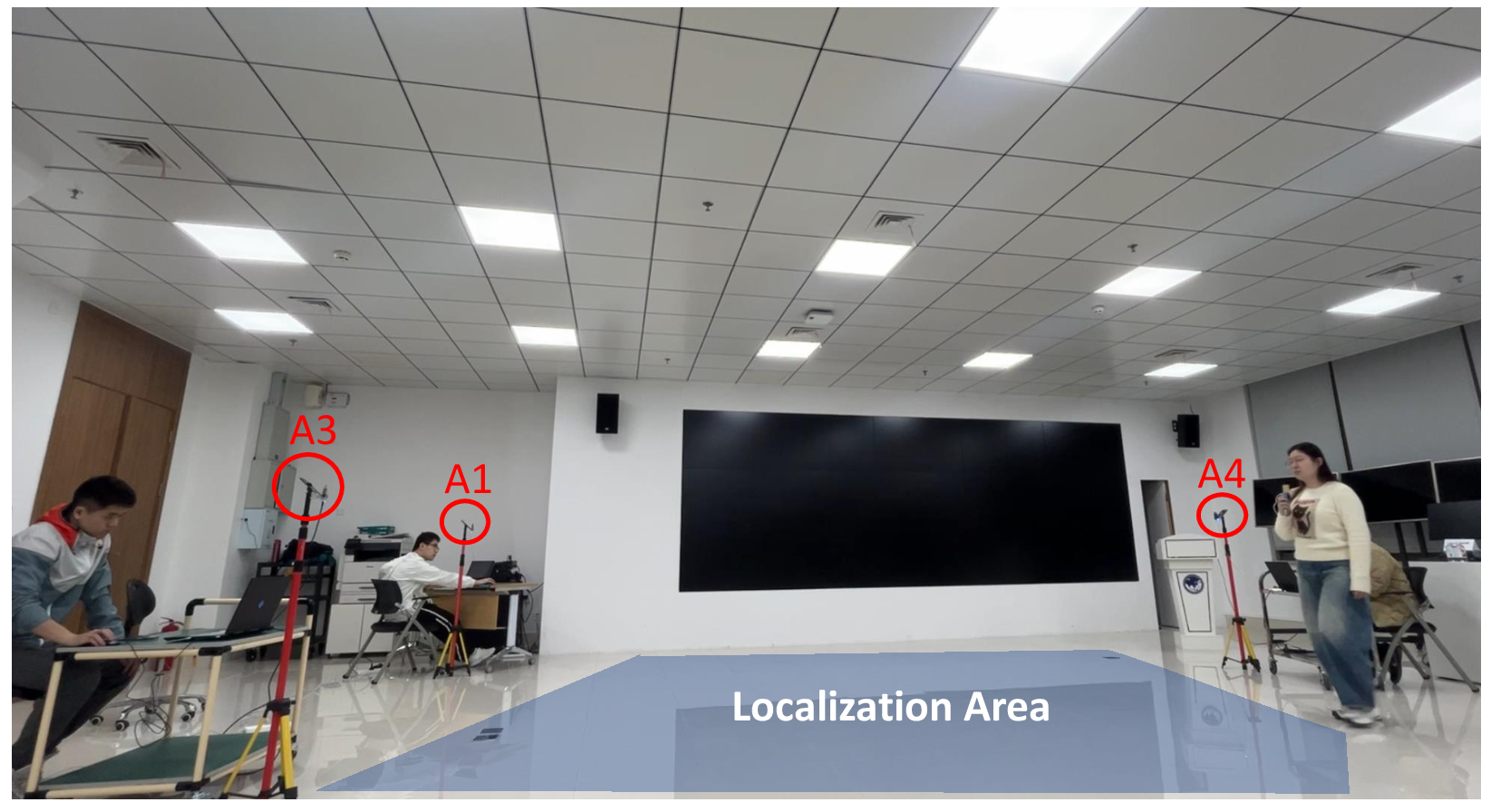}}  \quad\quad\quad
\subfigure[Experiment map and agent trajectory]{\label{FigExperimentEnviromentDiagram}\includegraphics[width = 0.8\columnwidth] {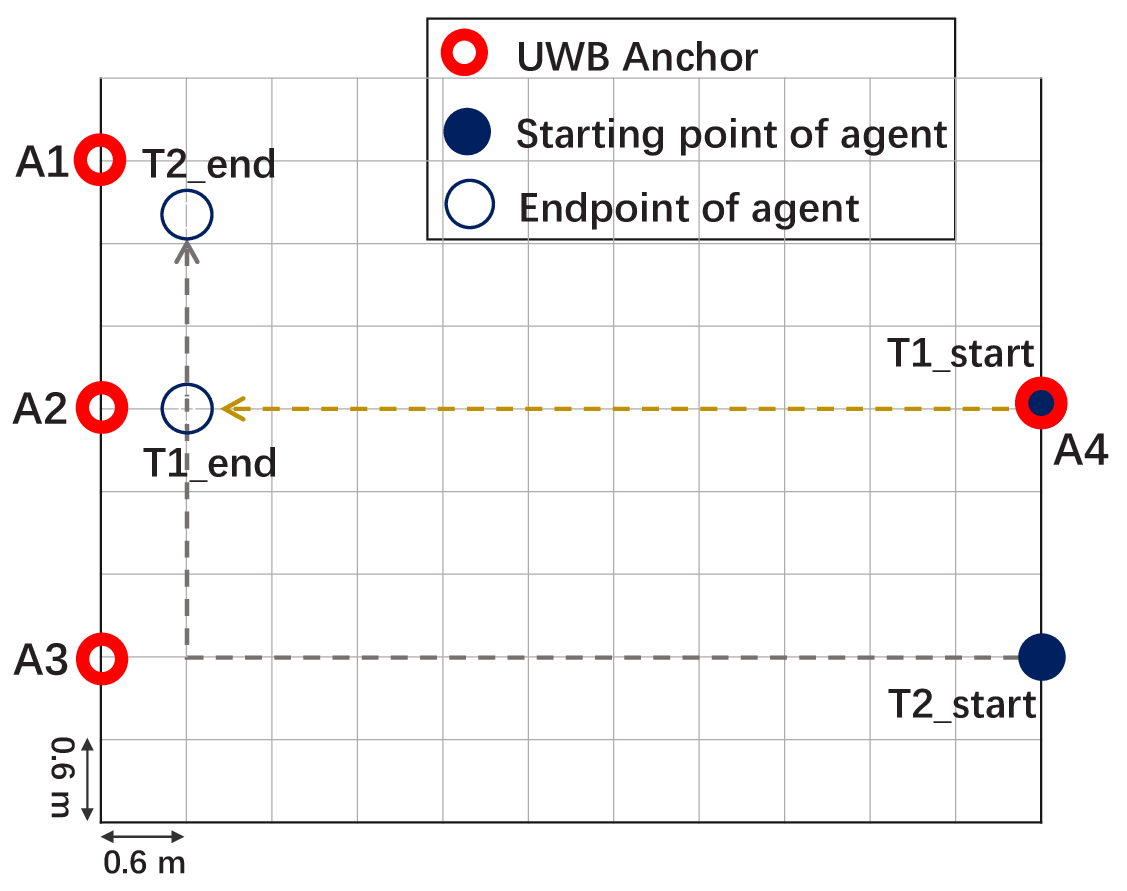}}
\caption{Experimental environments settings.}
\label{Fig Experimental environments}
\vspace{-10pt}
\end{figure*}

In this subsection, we describe the experimental setup for dynamic agent localization in a laboratory environment, employing a downlink localization framework with three UWB anchors deployed in each scenario, as illustrated in Fig.~\ref{Fig Experimental environments}. The UWB anchor transmit UWB signals into the air,  which are then received by the agent within the designated localization area. It is important to note that there are clock offsets and drifts between the anchors and the agent as illustrated as Fig.~\ref{FigClock}. The parameters of the UWB model and environment settings are detailed in Table~\ref{Tab_UWBModel}.

\begin{table}  [t]
 \renewcommand{\arraystretch}{1.2}
 \centering
 \setlength{\belowcaptionskip}{5pt}
 \caption{Experiment measurement parameters}    \label{Tab_UWBModel}
 \begin{tabular}{lll}
  \toprule
  & Description                            & Value              \\
  \midrule
  & Coordinates of Anchor 1 (m)            & $(0, 3.6)$         \\
  & Coordinates of Anchor 2 (m)            & $(0, 1.8)$         \\
  & Coordinates of Anchor 3 (m)            & $(0, 0) $          \\
  & Coordinates of Anchor 4 (m)            & $(7.2, 1.8)$       \\
  & Output signal sampling processing (GHz) & $1$               \\
  & Carrier frequency (GHz)                & $7.9872$           \\
  & Measurement frequency (Hz)             & $10$               \\
  & Signal bandwidth (MHz)                 & $500$              \\
  \bottomrule
 \end{tabular}
\end{table}

\subsubsection{Downlink TDOA localization using measurements from 3 Anchors \bf{(Case 1)}}    \label{sectionCase1}

\begin{figure} [t]
\centering
\subfigure[Clock calibration]{\label{FigClock calibration1}\includegraphics[width = 1\columnwidth] {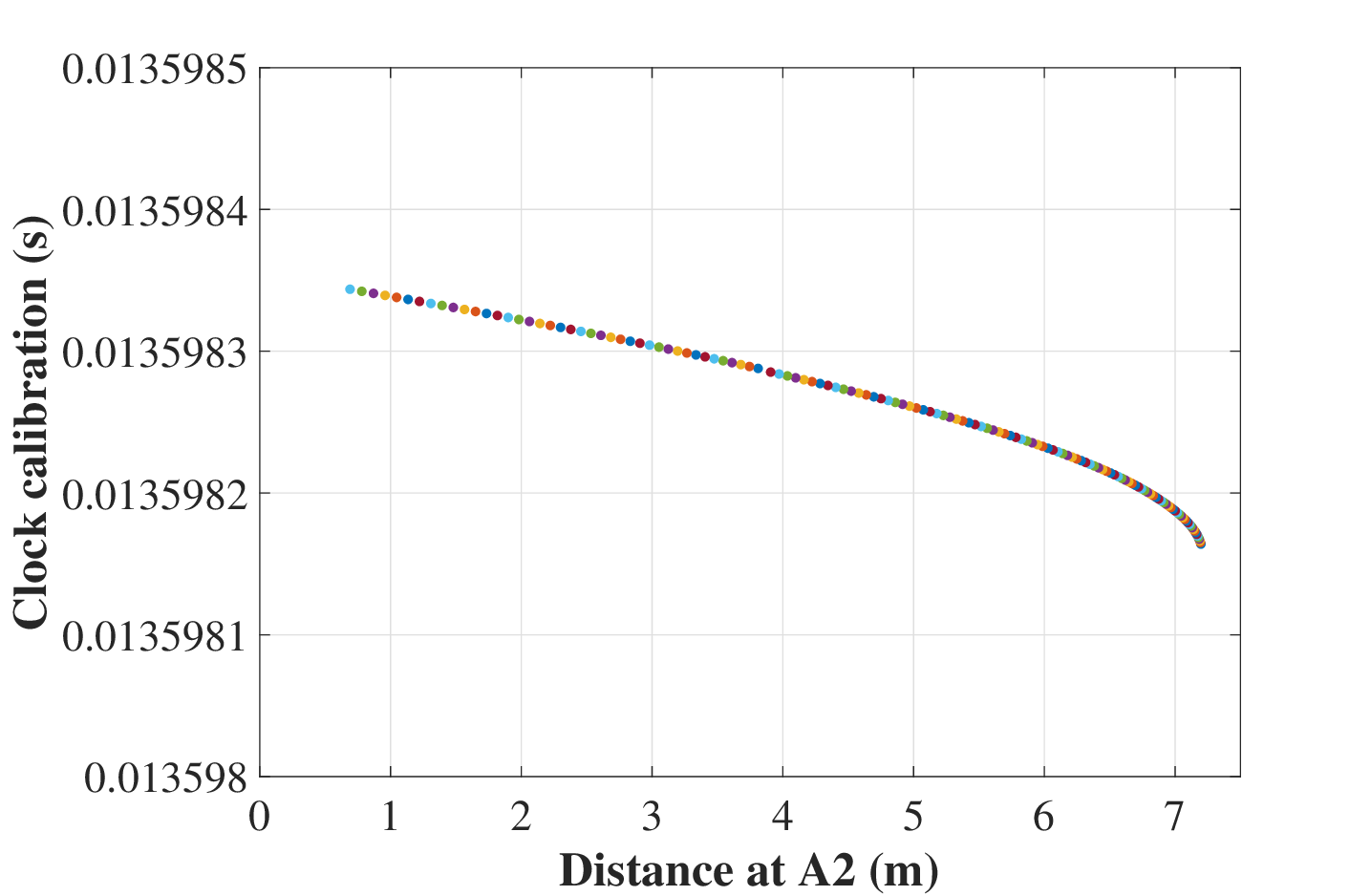}}  \quad\quad\quad\quad
\subfigure[Experiment map and agent trajectory]{\label{FigExperiment map1}\includegraphics[width = 1\columnwidth] {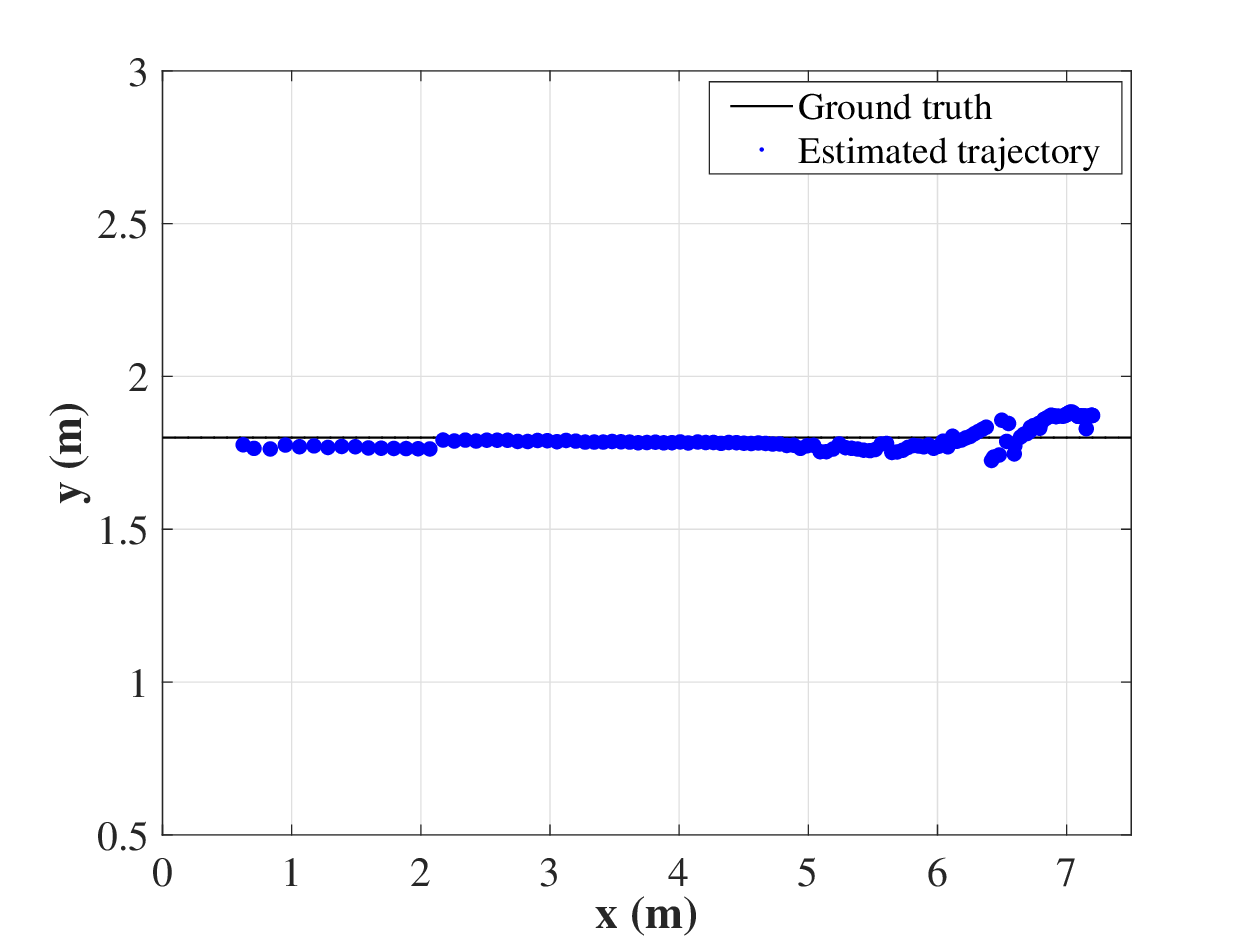}}
\caption{Clock calibration and localization of Case 1.}
\label{FigLocalizationResults1}
\end{figure}

\begin{figure} [t]
\centerline{\includegraphics[width = 1\columnwidth]{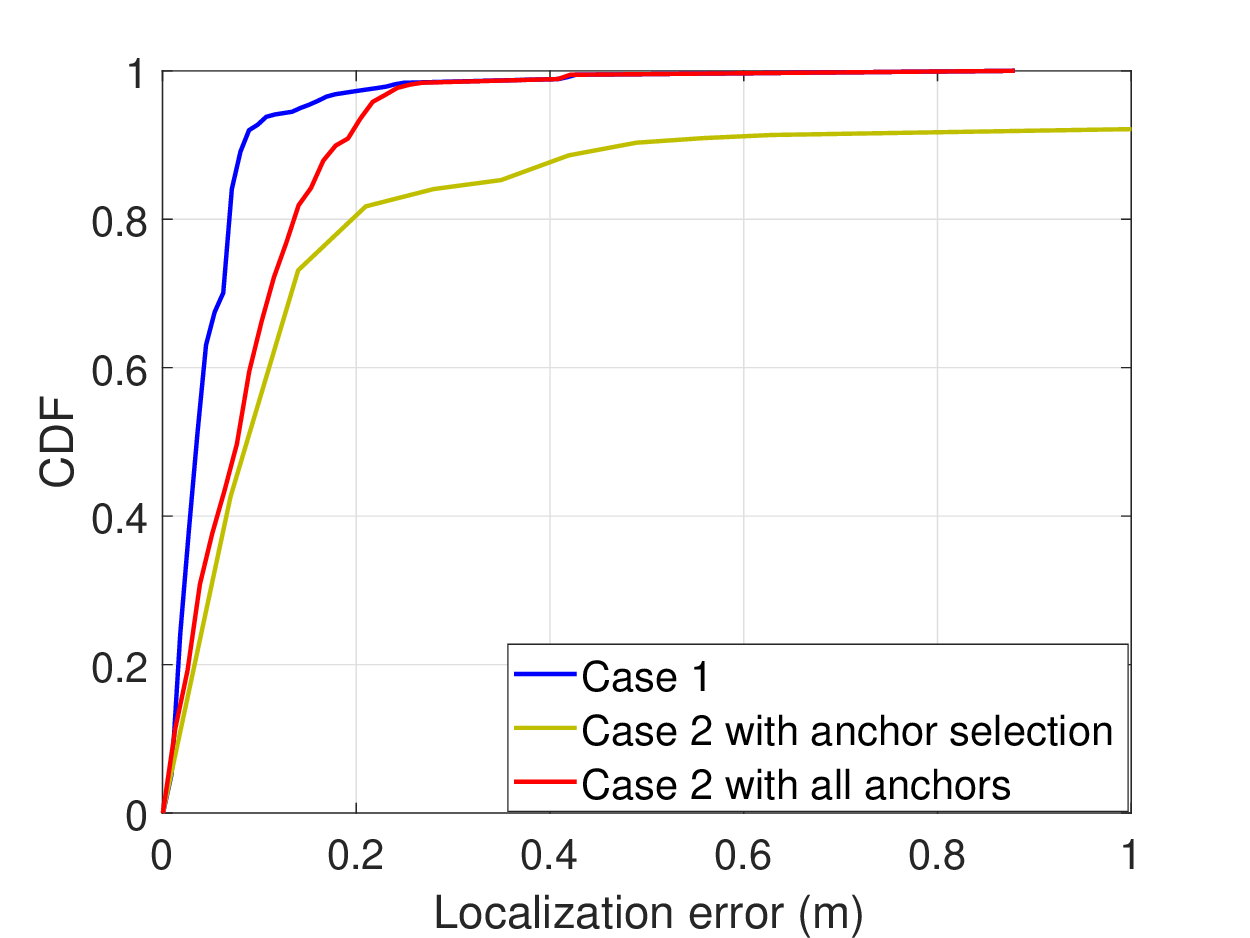}}
\caption{Localization results.}
\label{FigCDF}
\end{figure}

In this scenario, three anchors (A1, A2, and A3) are deployed for the localization of Agent 1, with the localization process spanning from T1\_start to T1\_end, as illustrated in Fig.~\ref{FigExperimentEnviromentDiagram}. Assuming the initial position of the agent is known, the clock offset and drift can be estimated. During the agent's movement, the range between the anchor and the agent is estimated following clock calibration. The clock-calibrated values and the estimated trajectory of the Agent 1 are depicted in Fig.~\ref{FigClock calibration1} and Fig.~\ref{FigExperiment map1}, respectively. It is evident that the calibrated values exhibit variations due to the presence of clock drift. As shown by the Case 1 curve in Fig.~\ref{FigCDF}, the localization results indicate that the probability of the localization error being less than 0.1 m exceeds 90\%.

\subsubsection{Downlink TDOA localization with measurements from two selected anchors \bf{(Case 2)}}    \label{sectionCase1}

\begin{figure} [t]
\centering
\subfigure[Clock calibration]{\label{FigClock calibration2}\includegraphics[width = 1\columnwidth] {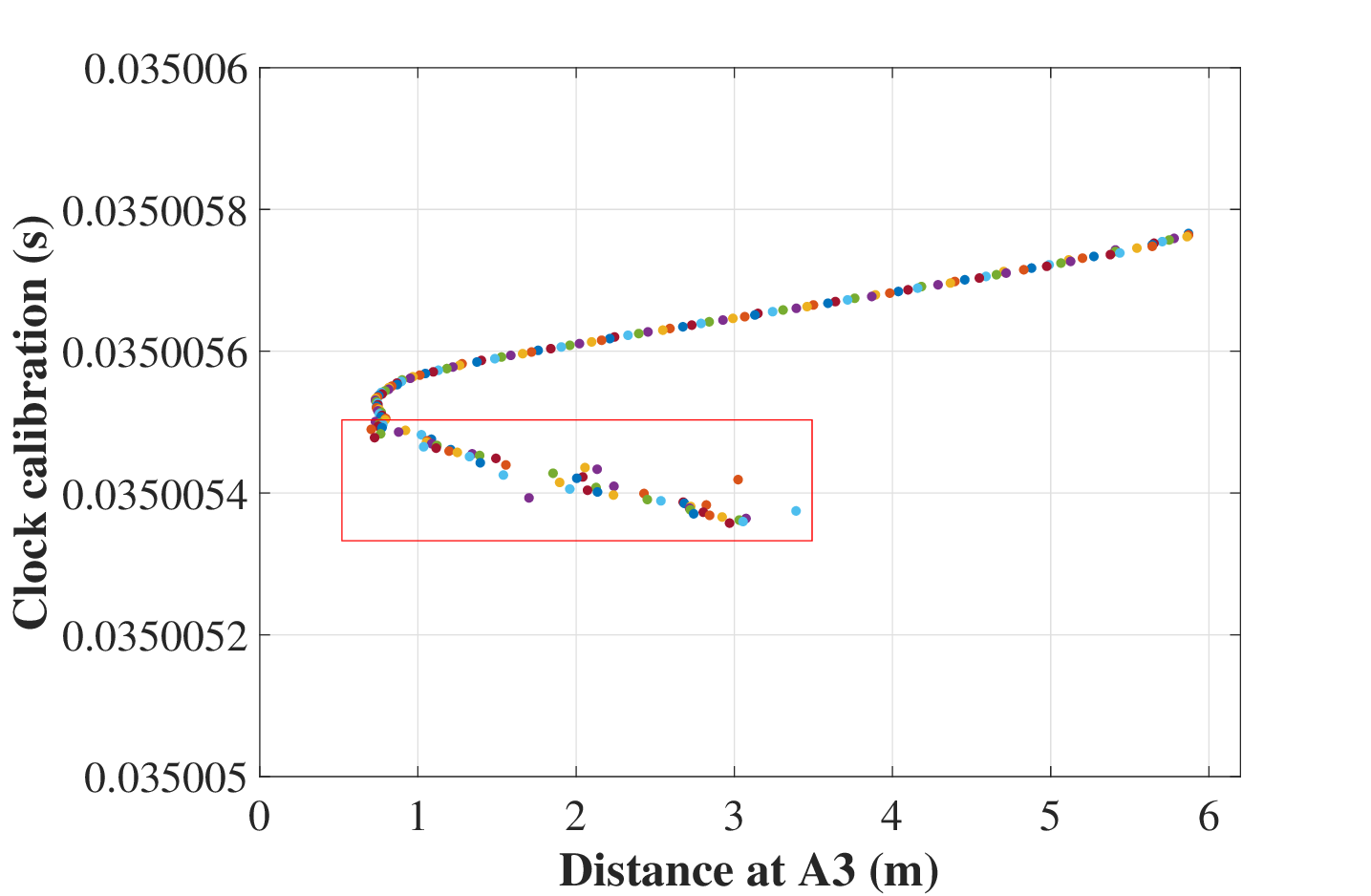}}  \quad\quad\quad\quad
\subfigure[Experiment map and agent trajectory]{\label{FigExperiment map2}\includegraphics[width = 1\columnwidth] {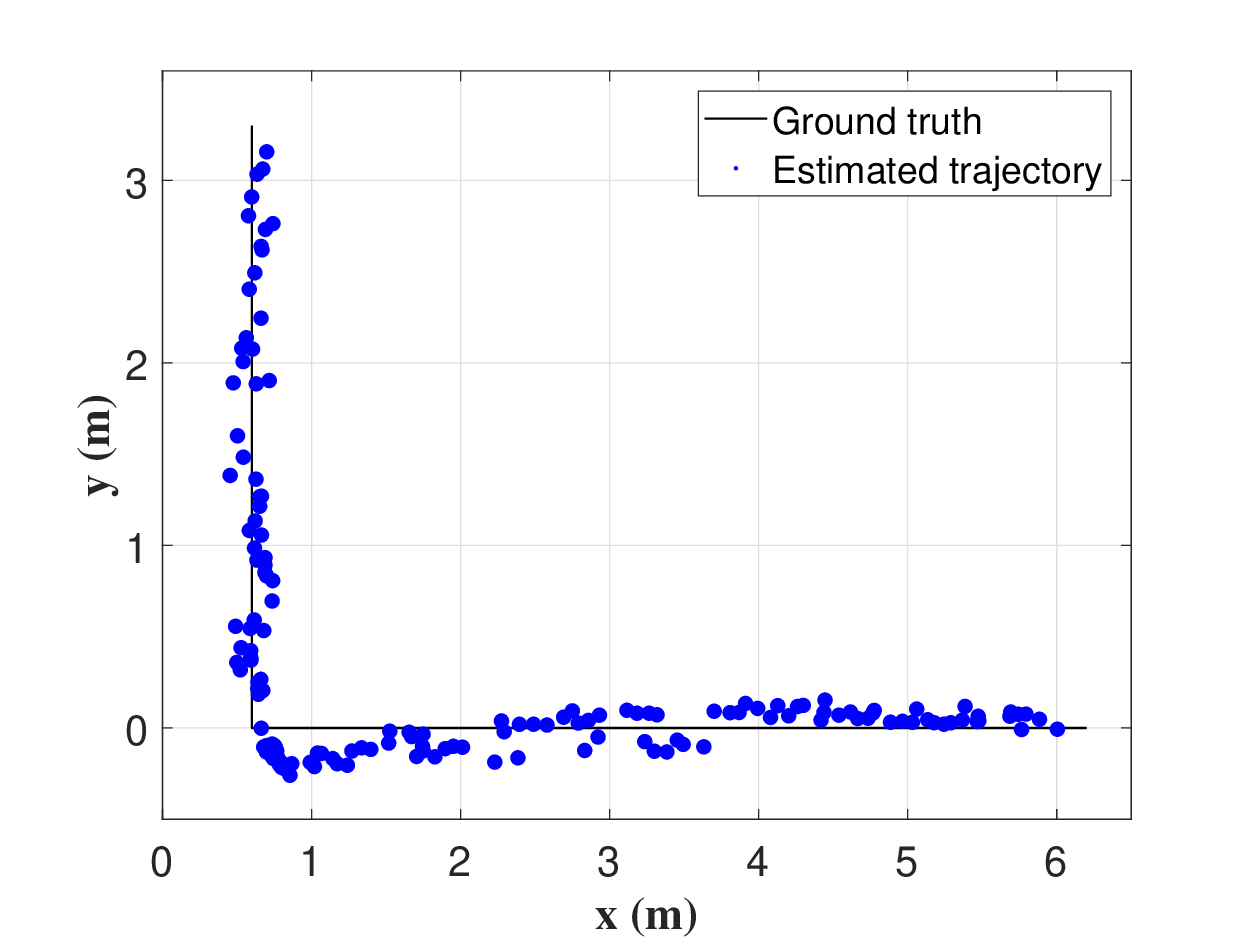}}
\caption{Clock calibration and localization of Case 2.}
\label{FigLocalizationResults2}
\end{figure}

In this case, three anchors (A1, A3, and A4) are deployed to localize the agent during its movement from T2\_start to T2\_end, as shown in Fig.~\ref{FigExperimentEnviromentDiagram}. During the agent's movement, the line-of-sight (LOS) path to a specific anchor may occasionally be obstructed, leading to significant measurement errors during the measurement process. To address this issue, we apply the confidence principle introduced in Sec.~\ref{sec_soft}, selecting the two anchors with the highest measurement confidence from the three available anchors to perform agent localization. The clock-calibrated values at A1 and the estimated trajectory of Agent 2 are presented in Fig.~\ref{FigClock calibration2} and Fig.~\ref{FigExperiment map2}, respectively. It is observed that the clock-calibrated results within the red box exhibit irregular behavior, which is attributed to the large range measurement error at A3 after the clock calibration.

Fig.~\ref{FigCDF} presents the localization results for Case 2, comparing scenarios using all anchors and selected anchors. The results demonstrate that the localization performance with two selected anchors achieves a cumulative distribution function (CDF) of 90\% for a root mean square error (RMSE) below 0.2 meters. In contrast, when three anchors are used but include low-confidence measurement distances, the CDF for RMSE $<$ 0.2 meters drops to 80\%. The optimal localization performance is observed when three anchors are utilized without low-confidence measurement distances, achieving a CDF of 90\% for RMSE $<$ 0.1 meters.

\section{Conclusions}       \label{section Conculsion }

This paper presents the design and implementation of an asynchronous IR-UWB multi-access ICL system. In particular, we have proposed a differential sequential detection strategy embedded within a confidence metric-based soft information framework for primary data demodulation. A theoretical model for SFD detection performance is established to support asynchronous system operation. During the initial static phase, clock drift is estimated using a WLS algorithm informed by the confidence of detected symbols. Subsequently, reliable data demodulation and localization are achieved within an appropriate confidence metric range.
Key findings include: (i) Increasing the transmitted signal power or the repetition count of a single pulse improves SFD detection performance, with maximum detection probability observed when symbol energy is uniform. (ii) A confidence threshold in the range [0.9, 0.95] significantly enhances clock drift estimation, achieving an accuracy of 0.1 $\emph{ppm}$. Within the confidence metric range [0.92, 0.994], both data demodulation and localization exhibit near-optimal performance in terms of minimal BER and RMSE. (iii) For fixed symbol energy, increasing the bit repetition count is more effective than increasing transmitted power in reducing communication BER and positioning RMSE. (iv) A real-world UWB downlink TDOA-based localization system is designed and evaluated, demonstrating a localization accuracy of 10 cm at a 90\% confidence level based on practical measurements. This work provides a robust framework for asynchronous downlink TDOA ICL systems, offering valuable insights for practical implementation.

\appendix

\subsection{Proof of Proposition \ref{ProCFAR}}   \label{AppendixProCFAR}

The decision variable at time $T$ after the MF of the received signals can be further expressed as
\begin{equation}   \label{eq discretizedSignal}
\begin{aligned}
& {\gamma_{\text {MF}}^{(mn)}}({T})
= \int_{0}^{{N_\text{r}}{T_f}} {{R^{(m)}}(t){\cal T}^{(n)}_{\kappa}(t)}dt  \\
&  = {\alpha^{(mn)}} \! \int_{0}^{{N_\text{r}}{T_f}} \!\! {\big({{s}}(t)* h(t)\big){\cal T}^{(n)}_{\kappa}(t)}dt
+ \! \int_{0}^{{N_\text{r}}{T_f}} \!\!\! {{z}(t){\cal T}^{(n)}_{\kappa}(t)}dt.
\end{aligned}
\end{equation}
Accordingly, the expectation and variance of ${\gamma_{\text {MF}}^{(mn)}}$ is
\begin{equation}
{\mu_{{\gamma_{\text {MF}}}}} = {\alpha^{(mn)}}\int_{0}^{{N_\text{r}}{T_f}} {\big({{s}}(t)* h(t)\big){\cal T}^{(n)}_{\kappa}(t)}dt,
\end{equation}
\begin{equation}
\sigma _{{\gamma_{\text {MF}}}}^2 = \int_{0}^{{N_\text{r}}{T_f}} {\big({\cal T}^{(n)}_{\kappa}(t)\big)^2}{\sigma _z^2}dt.
\end{equation}

A hypothesis testing model can be established by the MF decision variable as follows:
\begin{equation}
  \begin{split}
  & {H_0}:{\gamma_{\text {MF}}^{(mn)}} \sim N\big({\mu _0},\sigma _0^2 \big),       \\
  & {H_1}:{\gamma_{\text {MF}}^{(mn)}} \sim N\big({\mu _1},\sigma _1^2 \big),
  \end{split}
\end{equation}
where ${H_0}$ and $H_1$ indicate there is only noise, or signal with noise within one PRI, respectively.
\begin{equation}
  \begin{split}
  & {\mu _0} = 0, \;\;\;\;\;\;\ \sigma _0^2{\rm{\; = \;}}\sigma _{{\gamma_{\text {MF}}}}^2
             + {N_{\text r}}{\sigma _{\text {a}}^2}\sum\limits_{n = 2}^{N_{\text a}}A^{(mn)},       \\
  & {\mu _1} = {\mu _{{\gamma_{\text {MF}}}}},  \:\:\:  \sigma _1^2 = \sigma _{{\gamma_{\text {MF}}}}^2
             + {N_{\text r}}{\sigma _{\text {a}}^2}\sum\limits_{n = 2}^{N_{\text a}}A^{(mn)},
  \end{split}
\end{equation}
where the term ${N_{\text r}}{\sigma _{\text {a}}^2}\sum\limits_{n = 2}^{N_{\text a}}A^{(mn)}$ represents the variance caused by multi-anchor interference \cite{THwin},${\sigma _{\text {a}}^2} = {1 \over T_f}{\int_{0}^{T_f}\big[ {\int_{0}^{T_f}}w(t-\tau)w(t)dt \big]^2 d\tau}$,
$A^{(mn)} = {{\alpha^{(mn)}}\alpha^{{(mn)}}_{\text h}}\sqrt{E_{\text {tb}}}$ represents the total channel attenuation, $\alpha^{{mn}}_{\text h}$ represents the small-scale fading included in $h(t)$.

The false alarm probability (FAP) is expressed as
\begin{equation}
\begin{aligned}
{P_{\text {FA}}} & \: {\overset{\triangle} =} \: P\big( {{H_1}{\rm{ : }}{H_0}} \big)  \\
& = Q\Big({{\gamma  - {\mu _0}} \over {\sigma _0^{}}} \Big),
\end{aligned}
\end{equation}
where $Q(\cdot)$ is the right tail function of the standard normal distribution \cite{Qfunction}. According to the classical constant false alarm detection (CFAR) criterion, the threshold of pulse
detection is
\begin{equation}
\begin{aligned}   \label{eq thred}
\gamma & = {Q^{{\rm{ - }}1}}\left( {{P_{\text {FA}}}} \right) \sigma _0           \\
   & = {Q^{{\rm{ - }}1}}\left( {{P_{\text {FA}}}} \right)
   \sqrt {\int_{0}^{{N_\text{r}}{T_f}} {\big({\cal T}^{(n)}_{\kappa}(t)\big)^2}{\sigma _z^2}dt}.
\end{aligned}
\end{equation}

Accordingly, the detection probability of the SFD is defined as
\begin{equation}
\small
\begin{aligned}       \label{eq PD}
  {P_\text{D}} & \! = \! P\left( {{H_1}{\rm{ : }}{H_1}} \right)  \\
          & \! = Q{\Big({{\gamma  - {\mu _1}} \over {\sigma _1^{}}}\Big)} \\
          & \! = \! Q \! \Bigg(\! {Q^{{\rm{ - }}1}} \! \left( {{P_\text{FA}}} \right)\! - \!
          {{\alpha^{(mn)}}\int_{0}^{{N_\text{r}}{T_f}} {\big({{S}}(t)* h(t)\big){\cal T}^{(n)}_{\kappa}(t)}dt
           \over {\sqrt{ {\int_{0}^{{N_\text{r}}{T_f}} {\big({\cal T}^{(n)}_{\kappa}(t)\big)^2}{\sigma _z^2}dt}
           + {N_{\text r}}{\sigma_{\text {a}}^2}\sum\limits_{n = 2}^{N_{\text a}}A^{(mn)}}}}  \! \Bigg).
\end{aligned}
\end{equation}

\bibliographystyle{IEEEtran}
\bibliography{reference}
\end{document}